\documentstyle[12pt]{article}
\newcommand{\beq}{\begin{equation}}
\newcommand{\eeq}{\end{equation}}
\newcommand{\beqn}{\begin{eqnarray}}
\newcommand{\eeqn}{\end{eqnarray}}

\newcommand{\be}{\begin{equation}}
\newcommand{\ee}{\end{equation}}
\newcommand{\ba}{\begin{array}}
\newcommand{\ea}{\end{array}}
\newcommand{\bea}{\begin{eqnarray}}
\newcommand{\eea}{\end{eqnarray}}
\newcommand{\stackM}{\stackrel{\scriptstyle >}{{ }_{\sim}}}
\newcommand{\stackm}{\stackrel{\scriptstyle <}{{ }_{\sim}}}

\newcommand{\GeV}{\mbox{ GeV}}

\newcommand{\st}{{\tilde t}}
\newcommand{\sbt}{{\tilde b}}

\newcommand{\sbp}{{\tilde{b}_1}}
\newcommand{\stp}{{\tilde{t}_1}}
\newcommand{\stau}{{\tilde{\tau}}}
\newcommand{\sneut}{{\tilde {\nu}}}
\newcommand{\sg}{{\tilde g}}
\newcommand{\hplus}{{H^+}}

\newcommand{\cplus}{{\chi^+_1}}
\newcommand{\cmin}{{\chi^-_1}}
\newcommand{\neut}{{\chi^0_1}}

\newcommand{\pl}{P_L}
\newcommand{\pr}{P_R}

\newcommand{\slas}[1]{\rlap/ #1}

\newcommand{\tw}{\tan \theta_W}

\begin{document}

\thispagestyle{empty}
\def\pubnum{389}
\def\data{March, 1996}


\def\UUAABB{ \hfill\hbox{
    \vrule height0pt width2.5in
    \vbox{\hbox{\rm 
     UAB-FT-\pubnum
    }\break\hbox{\data\hfill}
     \break\hbox{hep-ph/9603441\hfill} 
    \hrule height2.7cm width0pt}
   }}   
\UUAABB
\vspace{3cm}
\begin{center}
\begin{large}
\begin{bf}
SUPERSYMMETRIC THREE-BODY DECAYS OF THE TOP QUARK IN THE MSSM\\
\end{bf}
\end{large}
\vspace{1cm}
Jaume GUASCH, Joan SOL\`A\\

\vspace{0.25cm} 
Grup de F\'{\i}sica Te\`orica\\ 
and\\ 
Institut de F\'\i sica d'Altes Energies\\ 
\vspace{0.25cm} 
Universitat Aut\`onoma de Barcelona\\
08193 Bellaterra (Barcelona), Catalonia, Spain\\
\end{center}
\vspace{0.3cm}
\hyphenation{super-symme-tric}
\hyphenation{com-pe-ti-ti-ve}
\begin{center}
{\bf ABSTRACT}
\end{center}
\begin{quotation}
\noindent
We survey all possible supersymmetric three-body decays of the top quark
in the framework of the MSSM and present detailed numerical analyses of the
most relevant cases. Although the two-body channels are generally dominant,
it is not inconceivable
that some or all of our most favourite two-body SUSY candidates could 
be suppressed. 
In this event there is still the possibility that some of the available
three-body SUSY modes might exhibit a substantial branching fraction and/or
carry exotic signatures that would facilitate their identification. 
Furthermore, in view of the projected inclusive measurement
of the top-quark width $\Gamma_t$ in future colliders, one should have at
one's disposal the full second order correction (electroweak and strong)
to the value of that parameter in the MSSM. 
Our analysis confirms that some supersymmetric three-body decays 
could be relevant and thus contribute to $\Gamma_t$  
at a level comparable to 
the largest one loop supersymmetric effects on two-body modes recently
computed in the literature.    
\end{quotation}
 
\baselineskip=6.5mm  


\newpage
\begin{Large}
 {\bf 1. Introduction}
\end{Large}
 \vspace{0.5cm}

The era of the top quark has just begun\,\cite{Tevatron}. To a large
extent we were prepared to enter the long-announced new epoch and in the  
meanwhile a tremendous amount of work has piled up on 
top-quark observables\footnote{See e.g. Refs.\,\cite{Bern,CPYuan} and references
therein.}. Yet the
extremely rich potential phenomenology and
the far-reaching consequences that top-quark dynamics may have on the final
picture that will emerge of the Standard Model (SM) of the electroweak
and strong interactions definitely demands a new fully fledged wave of
theoretical and experimental endeavor in Particle Physics.  

The SM has been a most succesful framework to describe the
phenomenology of the strong and electroweak interactions
for the last thirty years\,\cite{Hollik}.
The top quark itself stood, at a purely theoretical level --namely, on the grounds of
requiring internal consistency, such as gauge invariance and renormalizability--
as a firm prediction of the SM
since the very confirmation of the existence of the bottom quark and the measurement
of its weak isospin quantum numbers\,\cite{Zerwas}.
Nevertheless, despite all the successes there are still too many questions left
unanswered by the SM, especially in connection with the nature of the
spontaneous symmetry breaking mechanism (SSB) and the purported existence of  
the extremely elusive Higgs particle, whether realized as a truly
elementary particle or as an effective (composite) field. 
Due to the huge mass of the top quark,
one expects it to be the most preferential fermion to which the Higgs particle
should couple. Therefore, top physics is deemed to be the ideal environment
for Higgs phenomenology.

Lately the SM has been exposed to an escalate of experimental
information potentially challenging some of its predictions
to an unprecedented level. We are referring to the long-standing conundrum
involving the high
precision $Z$-boson observable $R_b$ and some related observables\,\cite{LEPEWWG}.
The issue about $R_b$ is specially offending, for it seems to consolidate
with time--the present
day discrepancy with the SM being at the $3.4\,\sigma$ level\,\cite{Hildreth}.
Whether this anomaly is linked to an incomplete understanding of the experimental
analysis of $Z$ decays into
b-quarks, or it can be licitly associated with some sort of
physical effects beyond the SM,  has not yet been
established and at present there is a lot of controversy about it\,\cite{Hildreth}.
Be as it may, from the theoretical point of view one is tempted to  
believe that physics might be taking a definite trend beyond the SM.
One possibility is to look at the predictions of the supersymmetric (SUSY)
extension of the SM. 
In this paper we take our chances in favour of the 
elementary Higgs particle interpretation of the SSB and we adhere to 
the supersymmetric extension of the SM, more specifically to the
Minimal Supersymmetric Standard Model (MSSM)\,\cite{MSSM}. 
In particular, there is in the literature
quite a big amount of early\,\cite{RbEarly} as well as of very recent SUSY work on
$R_b$\,\cite{Kaneboys}-\cite{ELN}. In some of these works it is
shown that the discrepancies, although they cannot be fully accounted for,
at least they can be significantly weakened under suitable
conditions\,\cite{GJS2,GS1,GS2}.

In view of the new wave of SUSY potentialities, 
it is natural to study all possible phenomenological
consequences that may be expected on supersymmetric top-quark physics; after all,
the interactions between the top-quark and Higgs sector are doubled
in a SUSY framework and one may hope
that top-quark physics can be a window to both Higgs bosons and 
supersymmetric particles.    
As a first step in this direction one would like to assess the 
importance of real and virtual SUSY effects on top-quark decays.
Here, too, a fairly respectable amount of work is scattered over the
literature\footnote{For a review of some of these
results, see Refs.\cite{JSAmes,JSMoriond}.}:

i) Supersymmetric two-body decays of the top quark have been described
at the tree level e.g. in
Refs.\cite{Bern}, \cite{Hunter}-\cite{Cuypers}; 

ii) Supersymmetric Higgs corrections to the conventional top-quark decay mode
$t\rightarrow W^+\,b$ have been computed in Ref.\cite{GHDH};

iii) Supersymmetric electroweak quantum effects on $t\rightarrow W^+\,b$
mediated by the roster of genuine supersymmetric particles, such as
sleptons, squarks, charginos and neutralinos, have been accounted for in
Refs.\cite{GJSH,YangLi};

iv) Supersymmetric QCD corrections to $t\rightarrow W^+\,b$ have also been
studied in detail in Ref.\cite{DHJJS};

v) Supersymmetric QCD effects on the charged Higgs decay
of the top quark, $t\rightarrow H^+\,b$, are generally relevant
and can be spectacularly large in favourable regions
of the MSSM parameter space\,\cite{GJS,LiYangKo};

vi) The full plethora of electroweak quantum effects on the unconventional mode
$t\rightarrow H^+\,b$ in the MSSM are also recently available and can be rather
significant\,\cite{JSMoriond,CGGJS}.

We see that on the theoretical side there is a large amount of work ready to
be used by experimentalists. Now,
what about the prospects for an experimental measurement of these effects?.  
On the one hand, the measurement of $\Gamma(t\rightarrow W^+\,b)$ will be carried out
at the Tevatron at a level of $\sim 10\%$ and will be further reduced
at the LHC.
In this respect we remind that the top-quark phenomenology
is expected to significantly
improve at the Tevatron\,\cite{CPYuan,Willenbrock} on the basis of a projected
ten-fold increase of the luminosity via the Main Injector and Recycler facilities,
together with a $\sim 35\%$ increase of the production cross-section at a $2\,TeV$
running energy (Run II), as compared to the fruitful $1.8\,TeV$ past run (Run I). 
However, in a hadron machine one aims more at a measurement of specific top-quark
production vertices, which are obviously related to the corresponding
top-quark partial decay widths.
For instance, one of the main goals at the Tevatron for Run II is the measurement
of the single top-quark production
cross-section\,\cite{Willenbrock,Heinson}, which gives essential information
on the SM vertex $t\,b\,W$ and, a fortiori, on the value of the CKM matrix element
$V_{tb}$. In the absence of new physics this measurement is equivalent to a 
determination of the top quark width.
However, in the presence of new interactions beyond the SM, such as e.g.
SUSY interactions, one may expect significant changes in the prediction
for the cross-section which can be related to the hadronic partial
widths of the Higgs bosons of the MSSM\,\cite{JimSol,CJS,Bartl}.

On the other hand, from the point of view of an {\sl inclusive} model-independent
measurement of  
the {\sl total} top-quark width, $\Gamma_t$, the future $e^+\,e^-$ supercollider 
should be a better suited machine. In an inclusive measurement,
all possible non-SM effects would appear on top
of the corresponding SM effects already computed in the
literature\,\cite{TopSM}. 
As shown in Ref.\cite{Fujii}, one expects to be able to measure the top-quark
width in $e^+\,e^-$ supercolliders at a level of $\sim 4\%$ on the basis of a
detailed analysis of both the top momentum distribution and the
resonance contributions to the forward-backward
asymmetry in the $t\bar{t}$ threshold region.
 
Clearly, for a consistent treatment of the corrections to the
observable $\Gamma_t$ at second order of perturbation theory (in the 
strong and electroweak gauge couplings) one should include the tree-level
contributions from all possible three-body decays
of the top quark in the MSSM. As it happens, the contribution of some
of these three-body decays turn out to be comparable to the largest SUSY quantum
effects on the two-body channels mentioned above. Furthermore, it could occur 
that our most cherished SUSY two-body decays are not kinematically allowed
or are significantly suppressed in some regions of parameter space.
Therefore, supersymmetric
three-body decays could be
relevant and a detailed study is in order.
Such a study is, to our knowledge, missing in the literature and
it is precisely the task that we have
undertaken in this article.
The paper is organized as follows. In Section 2 we present an overview of two-body
and three-body decays of the top quark in the MSSM. The Lagrangian interactions
relevant to these decays in the  MSSM  are given in Section 3. The numerical
analysis of the
various partial widths, with special emphasis on the dominant modes, is detailed in
Section 4. Finally, Section 5 is devoted to a discussion of the results, as well
as of the possible signatures for the favourite decays, and we 
deliver our conclusions.   
An appendix is provided at the end to display some lengthy formulae. 

\vspace{0.5cm}


\begin{Large}
 {\bf 2. Decays of the top quark in the MSSM}
\end{Large}
 \vspace{0.5cm}

The weighted average of the CDF and D$0$ determinations of the
top-quark mass reads as follows\,\cite{Tartarelli}:
\beq
m_t=175\pm 9\,\GeV\,.
\label{eq:CDFD0}
\eeq 
Due to the large mass of the top quark,
there is plenty of phase space available
for two-body and multibody decays. Within the minimal SM, the
leading standard decay of the top quark is the ``canonical'' decay 
\beq
t\rightarrow W^+\,b\,.
\label{eq:canonical}
\eeq
The associated partial width is given at the tree-level by
\beqn
\Gamma_{SM}\equiv\Gamma (t\rightarrow W^+\,b) & = &
\left({G_F\over 8\pi\sqrt{2}}\right){|V_{tb}|^2\over m_t} 
\ \lambda^{1/2} (1, {m_b^2\over m_t^2}, {M_W^2\over m_t^2})
\nonumber\\
& & \times [M_W^2(m_t^2+m_b^2)+ (m_t^2-m_b^2)^2-2M_W^4]\,,
\label{eq:treeW}
\eeqn  
where
\beq
\lambda^{1/2} (1, x^2, y^2)\equiv\sqrt{[1-(x+y)^2][1-(x-y)^2]}\,.
\eeq
Additional standard decays to other quarks are of course
possible but, contrary to the canonical decay (in which $V_{tb}\simeq 0.999$, for
three generations), they are CKM-suppressed.
Numerically, for $m_t=175\,\GeV$ (and $m_b=5\,\GeV$)  one has 
$\Gamma (t\rightarrow W^+\,b)\simeq 1.54\,\GeV$, which is
very big as compared to $\Lambda_{QCD}\stackm 300\,MeV$, and therefore the top quark
decay is basically a free-quark decay\,\cite{FaPesk}.
At this point we should recall that the latest measurements
of the canonical branching ratio at the Tevatron still
give room enough for top quark decays beyond the SM, namely they could reach
$BR(t\rightarrow {\rm ``new''})\simeq 40\%$\,\cite{Tartarelli}.

As for the SUSY two-body decays\,\cite{Bern},\cite{Hunter}-\cite{Cuypers},
the leading modes are the following. 
On the SUSY-QCD side the top quark can disintegrate, if there is phase
space enough, into the lightest
stop ($\tilde{t}_1$) and gluino
($\tilde{g}$), 
\beq
t\rightarrow \tilde{t}_1\,\tilde{g}\,,
\label{eq:tt1g}
\eeq
and on the SUSY electroweak side we have 
\beq
t\rightarrow \tilde{b}_1\,\chi^+_1\,,
\label{eq:tb1Psi1}
\eeq
and
\beq
t\rightarrow \tilde{t}_1\,\chi^0_1\,,
\label{eq:tt1Psi0}
\eeq
where $\tilde{b}_1$ stands for the lightest sbottom and $\chi^+_1$ ($\chi^0_1$)
for the lightest chargino (neutralino). In some cases the next-to-lightest neutralinos
can also enter and be relevant (see Sections 4-5).
Another conspicuous electroweak decay of the top quark in the MSSM involves 
a supersymmetric charged Higgs in the final state,
\beq
t\rightarrow H^+\,b\,.
\label{eq:tHb}
\eeq
This decay, if kinematically allowed, could be very promising
and it has recently been studied  
up to one-loop order in great detail in Ref.\cite{CGGJS}.
Numerically, all the two-body SUSY modes can be important as compared to the
canonical mode (\ref{eq:canonical}) in certain regions of the MSSM parameter
space. We shall compare their contribution with that of the leading three-body
decays in Section 4.

Concerning the three-body decays, there are the SM modes where the $W^+$ is a
virtual particle that subsequently goes into lepton or quark final states.
However, since the $W^+$ can be real, the two-body mode (\ref{eq:canonical}) 
followed by the decay of the $W^+$  as a real particle
is overwhelming in the SM. In fact, this is the way in which the 
top quark has been discovered\,\cite{Tevatron}. Therefore, we 
are only interested in three-body decays of the top quark
in the MSSM
other than the three-body SM decays. We shall christen these decays, the SUSY
three-body decays of the top quark. In contrast to the SM case, in the MSSM
not all of them need to be suppressed as compared to the two-body modes, as we
shall see.
In an extreme situation, the three-body decays could be the leading SUSY decays of
the top quark.
There are a fairly big number of them, but in the end only a few can be of some
relevance. In the following, and unless stated otherwise, we shall impose 
the following condition: the relevant
three-body decays under study are only those decays in which the virtual particle
is heavy enough that the corresponding SUSY two-body decay is kinematically forbidden.
Later on we shall
relax this condition in some especial cases. 

The following decays are essentially ruled out by our conditions or by 
direct phenomenological constraints:

\begin{enumerate}
\item\label{cu} $t \rightarrow \stp W^+ \cmin$: Forbidden by phase space:
\beq
\ba{lcccl}
m_\stp+M_\cmin& <& m_t-M_W &\simeq& 100 \GeV\\
m_\stp+M_\cmin &>& 120 \GeV\,.
\ea
\eeq
where the second relation reflects the recent LEP 130/140 bound on the lightest
chargino and sfermion masses\,\cite{LEP140}
\beq
m_{\tilde{f}_1},M_{\chi^{\pm}_1}\stackM 60\,\GeV\,,\ \ \ \
\label{eq:LEP140}
\eeq 
\item\label{cdos} $t \rightarrow \sbp \neut W^+$: In this case we require
\beq
\ba{c}
m_\sbp+M_\neut < m_t-M_W \simeq 100 \GeV \\
m_\stp+M_\neut > m_t \\
m_\sbp+M_\cplus > m_t\,, 
\ea
\eeq
so that 
\be
\ba{lcl}
M_\neut & < & 40 \GeV \\
M_\cplus & > &80 \GeV+M_\neut\,,
\ea
\label{eq:cdos}\ee
which cannot be fulfilled in the gaugino-higgsino
window $(\mu,M)$ --see Section 3-- of the MSSM
parameter space, as we have
checked.

\item\label{cset}\label{conze}$t \rightarrow \stp\,H^{\pm}\chi_1^{\mp}$:
Impossible in the MSSM
where  $m_\hplus\geq M_W$, and so we cannot fulfil the phase space constraint 
\beq
m_\hplus < m_t-m_\stp-M_\cmin \stackm 60 \GeV\,,
\eeq
not even in the presence of SUSY radiative corrections, which
would lower (only slightly)
the charged Higgs boson mass\,\cite{Higgsloop}.

\item\label{cvuit} $t \rightarrow \hplus \neut \sbp$,
\label{cnou} $t \rightarrow \hplus \sg \sbp$: For the first process, we have
the condition 
\beq
m_\hplus < m_t-M_\neut-m_\sbp\,,
\eeq
which forces the charged Higgs mass to be near the lowest limit allowed in the MSSM.
For the second, we may admit of both
light gluinos $m_{\tilde{g}}={\cal O}(1)\,\GeV$\,\cite{Clavelli} or 
heavy gluinos\,\cite{Abe},
\beq
m_{\tilde{g}}\geq 100\,\GeV\,.
\label{eq:hgluino}
\eeq
In the light gluino case, since 
\beq
m_\hplus < m_t-m_\sg-m_\sbp \stackm 120 \GeV\,,
\eeq
the two-body decay
$t \rightarrow \hplus b$ would already be allowed. In the heavy gluino scenario,
the decay at stake is excluded since it would enforce an 
unacceptably light charged Higgs:
\beq
m_\hplus < m_t-m_\sg-m_\sbp\stackm 20\,\GeV\,.
\eeq

\item\label{cdeu} $t \rightarrow \stp W^+ \cmin$: 
It would require
\beq
m_\stp+M_\cmin< m_t-M_W\simeq 100\,\GeV\,,
\eeq
which is ruled out by  eq. (\ref{eq:LEP140}).

\end{enumerate}

Among the decays in the complete list of SUSY three-body decays of the top-quark
in the MSSM which cannot be discarded by trivial arguments,
we remark the following:
\beqn
&{\bf I}.& t \rightarrow b\,\tau^+{\nu}_{\tau}\nonumber\\
&{\bf II}.& t \rightarrow h^0\, b\, W^+\nonumber\\
&{\bf III}.& t \rightarrow \tilde{b}_a\, W^+\, \sg\nonumber\\
&{\bf IV}.& t \rightarrow b\, \chi_{\alpha}^0\, \chi_i^+\nonumber\\
&{\bf V}.& t \rightarrow \tilde{t}_b\, \tilde{b}_a\, \bar b\nonumber\\
&{\bf VI}.& t \rightarrow \tilde{b}_a\,c\, \bar{\tilde{s}}_b\nonumber\\
&{\bf VII}.& t \rightarrow \tilde{b}_a\, \tau^+\, \bar{\tilde{\nu}_{\tau}}\nonumber\\
&{\bf VIII}.& t \rightarrow b\, \sg\, \chi_i^+\nonumber\\
&{\bf IX}.& t\rightarrow b\,\tilde{t}_b \, \bar{\tilde{b}}_a\nonumber\\
&{\bf X}.& t\rightarrow b\,\tilde{\tau}^+_a\, \tilde{\nu}_{\tau}\,,
\label{eq:processes}
\eeqn
where for each decay we sum over all sparticle indices allowed 
by phase space (see Section 4).
The decays quoted above do not exhaust the list of potential three-body 
modes, but the related
modes not included in the list are less significant. For instance,
decay I above proceeds both via a virtual $W^+$ and via a virtual charged
Higgs boson, $H^+$. The reason why this decay has been
singled out over similar decays involving light
quarks, e.g. $t \rightarrow b\,u\,\bar{d}$, is because for the
latters the charged Higgs couplings to light quarks  
are suppressed as compared to the coupling to a $\tau$-lepton. 
As another example, consider decay II.  It involves the lightest CP-even
SUSY Higgs boson, $h^0$\,\cite{Hunter}. Clearly, two related modes
are obtained by replacing $h^0$ with the heavy CP-even Higgs boson, $H^0$, or with 
the CP-odd Higgs boson, $A^0$. In the first case, the additional
decay will obviously be
suppressed by phase space; and in the second case, since the charged
Higgs $H^+$ is prescribed to be heavy enough in order to prevent the two-body mode
$t\rightarrow H^+\,b$ from occurring, it follows from the usual Higgs mass relations
in the MSSM\,\cite{Hunter} that the $A^0$ must be heavier than $h^0$,
and therefore the corresponding decay is phase-space obstructed.

The Feynman diagrams contributing to the various decays (\ref{eq:processes})
are given in Figs. 1-8. Each process has been 
thoroughly studied and the main numerical results
are provided in Section 4. The upshot of our analysis is that
there are a few selected decays in the list (\ref{eq:processes})
that could be of interest. As for the remaining decays,  
even though they cannot be dismissed by trivial arguments of the sort
used in the cases considered before, they eventually prove
to be irrelevant.  
\vspace{0.5cm}


\begin{Large}
 {\bf 3. Lagrangian interactions for top quark decays in the MSSM}
\end{Large}
 \vspace{0.5cm}

Although the Lagrangian of the MSSM is well-known\,\cite{MSSM}, it is always
useful to project explicitly the relevant pieces and to
cast them in a most suitable form for specific purposes. Even with
this arrangement, some care is to be exercised in actual calculations, due to
the Majorana nature of the neutralinos and the complicated mixing
among the various fields. 
We shall perform our calculations in a mass-eigenstate basis.
One goes from the weak-eigenstate basis to the mass-eigenstate basis via appropriate
unitary transformations. Two classes of SUSY particles enter our computations:
the fermionic partners of the weak-eigenstate gauge bosons and Higgs bosons
(called gauginos, $\tilde{B}$, $\tilde{W}$, and higgsinos, $\tilde{H}$, respectively)
and on the other hand the scalar partners of quarks
and leptons
(called squarks, $\tilde{q}$,  and sleptons, $\tilde{l}$,
respectively, or  sfermions, $\tilde{f}$, 
generically). Among the gauginos we also have the strongly interacting
gluinos, $\sg$, which are the fermionic partners of the gluons.
 Within the context of the MSSM,
we need two Higgs superfield doublets with weak hypercharges $Y_{1,2}=\mp 1$.
The (neutral components of the) corresponding scalar Higgs doublets
give mass to the down (up) -like quarks
through the VEV $<H^0_1>=v_1$ ($<H^0_2>=v_2$). The ratio
\beq
\tan\beta={v_2\over v_1}\,,
\label{eq:v2v1}
\eeq
is a most relevant parameter throughout our analysis.

Due to the ``chiral'' L-R mixing between weak-eigenstate
sfermions of a given flavor,
\beq
\tilde{q'}_a=\{\tilde{q'}_1\equiv \tilde{q}_L,\,\, 
\tilde{q'}_2\equiv \tilde{q}_R\}\,,
\eeq  
there is a matrix relation with the corresponding
squark mass-eigenstates,
\beq
\tilde{q}_a=\{\tilde{q}_1, \tilde{q}_2\}\,.
\eeq 
If we neglect intergenerational mixing, that relation is given by
\begin{eqnarray}
\tilde{q'}_a&=&\sum_{b} R_{ab}^{(q)}\tilde{q}_b,\nonumber\\ 
R^{(q)}& =&\left(\begin{array}{cc}
\cos{\theta_q}  &  \sin{\theta_q} \\
-\sin{\theta_q} & \cos{\theta_q} 
\end{array} \right)\;\;\;\;\;\;
(q=t, b)\,,
\label{eq:rotation}
\end{eqnarray}
where we have singled out 
the third quark generation, $(t,b)$, as a  generical 
fermion-sfermion generation. 
The rotation matrices in (\ref{eq:rotation}) diagonalize 
stop and sbottom mass matrices: 
\begin{equation}
{\cal M}_{\tilde{t}}^2 =\left(\begin{array}{cc}
M_{\tilde{t}_L}^2+m_t^2+\cos{2\beta}({1\over 2}-
{2\over 3}\,s_W^2)\,M_Z^2 
 &  m_t\, M_{LR}^t\\
m_t\, M_{LR}^t &
M_{\tilde{t}_R}^2+m_t^2+{2\over 3}\,\cos{2\beta}\,s_W^2\,M_Z^2\,.  
\end{array} \right)\,,
\label{eq:stopmatrix}
\end{equation}
\begin{equation}
{\cal M}_{\tilde{b}}^2 =\left(\begin{array}{cc}
M_{\tilde{b}_L}^2+m_b^2+\cos{2\beta}(-{1\over 2}+
{1\over 3}\,s_W^2)\,M_Z^2 
 &  m_b\, M_{LR}^b\\
m_b\, M_{LR}^b &
M_{\tilde{b}_R}^2+m_b^2-{1\over 3}\,\cos{2\beta}\,s_W^2\,M_Z^2\,,  
\end{array} \right)\,,
\label{eq:sbottommatrix}
\end{equation}
with 
\beq
M_{LR}^t=A_t-\mu\cot\beta\,, \ \ \ \ M_{LR}^b=A_b-\mu\tan\beta\,, 
\eeq
$\mu$ being the SUSY Higgs mass parameter in the superpotential, 
$A_{t,b}$ are the trilinear soft SUSY-breaking parameters and the
$M_{{\tilde{q}}_{L,R}}$ are soft SUSY-breaking masses\,\cite{MSSM}.
By $SU(2)_L$-gauge invariance we must have $M_{\tilde{t}_L}=M_{\tilde{b}_L}$,
whereas $M_{{\tilde{t}}_R}$, $M_{{\tilde{b}}_R}$ are in general independent
parameters. With regard to supersymmetric fermionic partners, 
from the higgsinos
and the various gauginos we form the following three sets of two-component
Weyl spinors:
 \begin{equation}
\Gamma_i^{+} = \{-i\tilde{W}^{+}, \tilde{H_2^{+}}\}\ , \;\;\;
\Gamma_i^{-} = \{-i\tilde{W}^{-}, \tilde{H_1^{-}}\}\ ,
\end{equation}
\begin{equation}
\Gamma_{\alpha}^{0} =
\{-i\tilde{B^{0}}, -i\tilde{W_3^{0}}, \tilde{H_2^0}, \tilde{H_1^0}\}\ .  
\end{equation}
These states also get mixed up when the neutral Higgs fields acquire nonvanishing
VEV's. The ``ino'' mass Lagrangian reads
\begin{equation}
{\cal L}_M=-<\Gamma^{+}|{\cal M}|\Gamma^{-}>-{1\over2}
<\Gamma^0|{\cal M}^0|\Gamma^0>+h.c.\ ,
\label{eq:CNMM}
\end{equation}
where the charged and neutral gaugino-higgsino mass matrices ${\cal M},{\cal M}^0$
are also well-known; in our notation they 
are given explicitly in Ref.\cite{GJSH}, where we remark the presence of
the parameter $\mu$ introduced above and of the 
soft SUSY-breaking Majorana masses $M$ and $M'$, usually related as
$M'/ M=(5/3)\,\tan^2{\theta_{W}}$.
The corresponding four-component 
mass-eigenstate spinors, the
so-called charginos and neutralinos, are the 
following\footnote{We use
the notation of Ref.\cite{GJSH}, namely,
first Latin indices a,b,...=1,2 are reserved for sfermions,
middle Latin indices i,j,...=1,2
for charginos, and first Greek
indices $\alpha, \beta,...=1,2,...,4$ for neutralinos. }:
\begin{equation}
 \Psi_i^{+}= \left(\begin{array}{c}
U_{ij}\Gamma_j^{+} \\ V_{ij}^{*}\bar{\Gamma}_j^{-}
\end{array} \right)
 \; \;\; \;\;,\;\;\;\;\;
 \Psi_i^{-}= C\bar{\Psi_i}^{+T} =\left(\begin{array}{c}
V_{ij}\Gamma^{-}_j \\ U_{ij}^{*}\bar{\Gamma}_j^{+} 
\end{array} \right)\ ,  
\label{eq:cinos}
\end{equation}
and 
\begin{equation}
 \Psi_{\alpha}^0= \left(\begin{array}{c}
N_{\alpha,\beta}\Gamma_{\beta}^0 \\ 
N_{\alpha,\beta}^{*}\bar{\Gamma}_{\beta}^0
\end{array} \right) =  
C\bar{\Psi}_{\alpha}^{0T}\ .
\label{eq:ninos} 
\end{equation}
where the matrices $U,V,N$ are defined through
\begin{equation}
U^{*}{\cal M}V^{\dagger}=diag\{M_1,M_2\}\;\;\;\;\;,\;\;\;\;\; 
N^{*}{\cal M}^0N^{\dagger}=diag\{M_1^0,...M_4^0\}\ .
\end{equation}    

With the help of these matrices, the following interaction
vertices appear in the SUSY
three-body decays of the top quark
(after rewriting them in the mass-eigenstate basis for all sparticles):

\begin{itemize}
\item{fermion--sfermion--(chargino or neutralino): 
\be
\ba{lcl}
{\cal L}&=&-g {\st}_a^{ *}\, \bar \chi_i^-\left({A_{+}}_{ai}^{t}
 \epsilon_i \pl+{A_{-}}_{ai}^{t}\pr\right) b\\
~&~&-g {\sbt}_a^{ *} \,\bar \chi_i^+\left({{A_{+}}_{ai}}^b\pl+{A_{-}}_{ai}^{b}
\epsilon_i \pr\right) t\\
~&~&-\frac{g}{\sqrt{2}} {\st}_a^{ *}\, 
\bar \chi_\alpha^0\left({A_{+}}_{a\alpha}^{(0)t}\pl+{A_{-}}_{a\alpha}^{(0)t}
\epsilon_\alpha\pr\right) t\\
~&~&+\frac{g}{\sqrt{2}} {\sbt}_a^{ *}\, 
\bar \chi_\alpha^0\left({A_{+}}_{a\alpha}^{(0)b}\pl+
{A_{-}}_{a\alpha}^{(0)b}\epsilon_\alpha\pr\right) b\\
~&~&+\mbox{ h.c.}
\ea
\label{eq:Lqsqcn}
\ee
Here  $P_{L,R}=\frac{1}{2}\left(1\mp\gamma^5\right)$ are the chiral
projectors, and 
we have defined the following coupling matrices:
\beq
\ba{l}
\left\{ \ba{lcl}
{A_{+}}_{ai}^{t}&=& R_{1a}^{(t)*} U_{i1}^* - \lambda_t R_{2a}^{(t)*} U_{i2}^* \\
{A_{-}}_{ai}^{t}&=& - R_{1a}^{(t)*} \lambda_b V_{i2}\\
\ea\right.\\
\left\{ \ba{lcl}
{A_{+}}_{ai}^{b}&=& R_{1a}^{(b)*} V_{i1}^* - \lambda_b R_{2a}^{(b)*} V_{i2}^* \\
{A_{-}}_{ai}^{b}&=& - R_{1a}^{(b)*} \lambda_t U_{i2}\\
\ea\right.\\
\left\{ \ba{lcl}
{A_{+}}_{a\alpha}^{(0)t}&=& R_{1a}^{(t)*} \left( N_{\alpha 2}^*+Y^L
\tw N_{\alpha 1}^*\right)
                        +R_{2a}^{(t)*} \sqrt{2}\lambda_t N_{\alpha 3}^*\\
{A_{-}}_{a\alpha}^{(0)t}&=& R_{1a}^{(t)*} \sqrt{2} \lambda_t N_{\alpha 3}
                        +R_{2a}^{(t)*} Y_t^R \tw N_{\alpha 1}\\
\ea\right.\\
\left\{\ba{lcl}
{A_{+}}_{a\alpha}^{(0)b}&=& R_{1a}^{(b)*} \left( N_{\alpha 2}^*-Y^L
\tw N_{\alpha 1}^*\right)
                        -R_{2a}^{(b)*} \sqrt{2}\lambda_b N_{\alpha 4}^*\\
{A_{-}}_{a\alpha}^{(0)b}&=& -R_{1a}^{(b)*} \sqrt{2} \lambda_b N_{\alpha 4}
                        -R_{2a}^{(t)*} Y_b^R \tw N_{\alpha 1}\,,
\ea\right.
\ea\
\label{eq:AS}
\eeq
where $Y^L$ and  $Y_{t,b}^R$ are respectively the weak hypercharges of the 
left-handed $SU(2)_L$ doublet and right-handed singlet fermion-sfermion
partners within the chiral supermultiplet. 
The potentially significant Yukawa couplings are contained in the
following ratios with respect to the $SU(2)_L$ gauge coupling:
\beq
\lambda_t\equiv{h_t\over g}={m_t\over \sqrt{2}\,M_W\,\sin{\beta}}\;\;\;\;\;,
\;\;\;\;\;\lambda_b\equiv{h_b\over g}={m_b\over \sqrt{2}\,M_W\,\cos{\beta}}\,.
\label{eq:Yukawas} 
\eeq

In the previous formulae we have changed the chargino-neutralino basis from
$\{\Psi_i^{\pm}, \Psi_\alpha^0\}$ to $\{\chi_i^{\pm}, \chi_\alpha^0\}$ and in the latter
we have introduced the coefficients $\epsilon_i$ and $\epsilon_{\alpha}$.
The reason for this change of basis is because in the numerical analysis we use
real, instead of complex, diagonalization matrices
$U,V,N$ for charginos and neutralinos. Thus we have to compensate for the minus
signs that may appear in the list of mass eigenvalues by introducing the
$\epsilon$ parameters.
For instance, for chargino masses we set, for each $i=1,2$: 
\beq
M_i=\epsilon_i |M_i|\,.
\eeq
In this formalism the physical chargino, $\chi_i^{\pm}$, is not always the 
the mass-eigenstate spinor defined in
eq.(\ref{eq:cinos}), but
\beq
\chi_i^{\pm}=\left\{\ba{ll}
\Psi_i^{\pm}&\,\ {\rm if}\ \ \epsilon_i=1\\
\pm\gamma_5 \Psi_i^{\pm} &\,\ {\rm if}\ \ \epsilon_i=-1\,.
\ea\right.
\eeq
This is equivalent to replace
\beqn
\Psi_i^+&\rightarrow &\chi_i^+\equiv\left(\pr+\epsilon_i\pl\right) \Psi_i^+\nonumber\\
\Psi_i^-&\rightarrow &\chi_i^-\equiv\left(\pl+\epsilon_i\pr\right) \Psi_i^-
=C\bar{\chi_i}^{+T}\,.
\label{eq:defepsilon}
\eeqn
Indeed, with this proviso the new  kinetic Lagrangian for the chargino has 
always the correct sign:
\beq
{\cal L}_{K}=\bar \Psi_i^+ \left( i \slas{\partial} - M_i \right) \Psi_i^+ 
=\bar \chi_i^+\left( i \slas{\partial} - |M_i| \right) \chi_i^+\,.
\eeq
We proceed similarly with neutralinos, for each $\alpha=1,...,4$:
\beq
M_{\alpha}=\epsilon_{\alpha} |M_{\alpha}|\,,
\eeq
and
\be
\Psi^0_{\alpha}\rightarrow 
\chi^0_{\alpha}\equiv\left(\pr+\epsilon_{\alpha}\pl\right) \Psi^0_{\alpha}\,.
\label{eq:defepsilon2}
\eeq
Notice, however, that 
in the real formalism just sketched the physical neutralinos
are no longer Majorana particles in general (Cf. eq.(\ref{eq:ninos})), since
they satisfy
\be
\chi^{0 c}_\alpha={\cal C} \bar\chi^{0 T}_\alpha = \epsilon_\alpha \chi^0_\alpha\,.
\label{eq:psiconjugat}
\ee
Once these definitions have been made one has to propagate them carefully over all the
interaction terms and keep track of the $\epsilon$ parameters\,\footnote{This
procedure is equivalent to the one defined in Refs.\cite{Hunter,Gunion}.
We have corrected some sign errors and some missing $\epsilon$'s in the Feynman
rules given in these references.}. There are a few
more or less known subtleties related to Majorana particles
in connection with the $\epsilon$ formalism which are worth remembering. Thus e.g.
take a generical Lagrangian interaction involving a neutral Higgs and
two neutralinos:
\beq
\ba{lcl}
{\cal L}&=&g H^0 \bar \chi^0_\gamma 
\epsilon_\delta \Gamma_{\gamma \delta}\pl \chi_\delta^0+\mbox{ h.c.}\\
~&=& g H^0\bar \chi^0_\gamma
 \left( \epsilon_\delta \Gamma_{\gamma \delta}\pl
 + \epsilon_\gamma \Gamma_{\delta \gamma}^* \pr \right)\chi_\delta^0\,.
\label{eq:alphabeta}
\ea
\eeq
Since $\chi^0_{\alpha}$ can be created or destroyed by any of the
fermionic field operators
appearing in the Lagrangian, for a term with fixed indices $\alpha$ 
and $\beta$ on eq.(\ref{eq:alphabeta}) we have
\beq
\ba{lcl}
{\cal L}_{\alpha\beta}&=&g H^0 \bar \chi^0_\alpha\left( \epsilon_\beta
 \Gamma_{\alpha \beta}\pl+\epsilon_\alpha \Gamma_{\beta \alpha}^* \pr
  \right)\chi_\beta^0 \\
~&~& + g H^0 \bar \chi^0_\beta\left( \epsilon_\alpha 
\Gamma_{\beta \alpha}\pl+\epsilon_\beta
 \Gamma_{\alpha \beta}^* \pr  \right)\chi_\alpha^0\,. 
\label{twoterms}
\ea
\eeq
Therefore, using eq.(\ref{eq:psiconjugat}) and the usual properties of the 
charge conjugation matrix, we may rearrange the two terms in 
(\ref{twoterms}) as follows:
\beq
{\cal L}_{\alpha\beta}=g H^0 \bar \chi^0_\alpha\left(
 \epsilon_\beta (\Gamma_{\alpha \beta}+\Gamma_{\beta \alpha})\pl+
\epsilon_\alpha (\Gamma_{\beta \alpha}^*+\Gamma_{\alpha \beta}^*) \pr
\right)\chi_\beta^0\,. 
\eeq
  
Notice that, for virtual particles, the 
rule (\ref{eq:defepsilon}) just
entails the following replacement in the numerator of the chargino propagator: 
\beq
\slas{p}+|M_i| \longrightarrow \slas{p}+\epsilon_i |M_i| = \slas{p} +M_i\,,
\eeq
and similarly for neutralinos,
so that one can use real matrices $U,V,N$ together with positive or negative mass
eigenvalues, since the $\epsilon$'s just cancel out\,\cite{GJSH}.
For real sparticles in the
final state, a similar thing occurs for the square of the various amplitudes
involved in the top quark decay, but here one has to keep track of 
the $\epsilon$'s anyway
since they may play a crucial role in the computation
of the interference terms. In our formulae we shall nevertheless maintain the
original complex notation  
so that one can either use complex mixing matrices and forget
altogether about $\epsilon$'s or, alternatively,
to keep the  $\epsilon$'s and consider that
all the coupling matrices are real.

}
\item{quark--squark--gluino:
\be
{\cal L}=- \frac{g_s}{\sqrt{2}}
\tilde q_{a,k}^{ *}\, \bar\sg_r 
\left(\lambda^r\right)_{k,l} \left(R_{1a}^{(q)*} \pl - R_{2a}^{(q)*} \pr \right) q_l
+\mbox{ h.c.}
\label{eq:Lqsqglui}\ee
where $\lambda^r$ are the Gell-Mann matrices. This is just the SUSY-QCD Lagrangian
written in the squark mass-eigenstate basis.
}
\item squark--squark--Higgs:
\be
{\cal L}=-\frac{g}{\sqrt{2}M_W} \st_c^{ *}
 \sbt_d\, R_{ac}^{(t)*} R_{bd}^{(b)} g_{ab} \hplus
+\mbox{ h.c.}
\label{eq:hsfsf}\ee
where we have introduced the coupling matrix
\be
g_{ab}=\left(
\ba{cc}
M_W^2\sin{2\beta}-\left(m_b^2 \tan\beta+m_t^2\cot\beta\right) &
 - m_b\left(\mu+A_b \tan\beta\right) \\
-m_t\left(\mu+A_t\cot\beta\right)&-m_t m_b \left(\tan\beta+\cot\beta\right)
\ea
\right)\,.
\label{eq:gij}
\ee
\item chargino--neutralino--Higgs: 
\be
{\cal L}=-g H^+\, \bar \chi^+_i \left(\sin\beta\, Q_{\alpha i}^{ R *}
 \epsilon_\alpha \pl
+\cos\beta\, Q_{\alpha i}^{ L *}\epsilon_i \pr \right) \chi^0_\alpha+{\mbox{ h.c.}}\,,
\ee
with
\beq
\left\{
\ba{lcl}
Q_{\alpha i}^{ L}&=&U_{i1}^*N_{\alpha 3}^*+
\frac{1}{\sqrt{2}}\left(N_{\alpha 2}^*+\tan\theta_W N_{\alpha 1}^*\right)U_{i2}^*\\
Q_{\alpha i}^{ R}&=&V_{i1}N_{\alpha 4}-
\frac{1}{\sqrt{2}}\left(N_{\alpha 2}+\tan\theta_W N_{\alpha 1}\right)V_{i2}\,.
\ea
\right.
\eeq
\item Gauge interactions: In our calculation we only need the sparticle 
interactions with the $W^\pm$:
\begin{itemize}
\item quarks:
\be
{\cal L}=\frac{g}{\sqrt{2}}\, \bar t \gamma^\mu \pl b\, W_\mu^++\mbox{ h.c.}
\label{eq:Lqqw}
\label{eq:Wq}
\ee
\item squarks:
\be
{\cal L}=i \frac{g}{\sqrt{2}}
R_{1b}^{(t)*} R_{1a}^{(b)} W^+_\mu \st_b^{ *}\stackrel{\leftrightarrow}
{\partial^\mu}\sbt_a
+\mbox{ h.c.}
\label{eq:LsqsqW}
\ee
\item charginos and neutralinos:
\be
{\cal L}=g\, \bar\chi_\alpha^0 \gamma^\mu \left(C_{\alpha i}^L 
\epsilon_\alpha \epsilon_i\pl + C_{\alpha i}^R \pr
\right)\chi_i^+\, W^-_\mu+\mbox{ h.c.}
\label{eq:WCN}
\ee
\beq
\left\{
\ba{lcl}
C_{\alpha i}^L&=&\frac{1}{\sqrt{2}}N_{\alpha 3}U_{i 2}^*-N_{\alpha 2}U_{i 1}^*\\
C_{\alpha i}^R&=&-\frac{1}{\sqrt{2}}N_{\alpha 4}^* V_{i 2}-N_{\alpha 3}^* V_{i 1}\,.
\ea\right.
\eeq
\end{itemize}
\end{itemize}

There are of course some additional interactions that can be
involved in the SUSY 
three-body decays of the top quark, such as the 
various quark-quark-higgs vertices, and also the
vertices involving one gauge boson and two higgses or viceversa. 
However, they are pretty well-known and we shall not
quote them here\,\cite{Hunter}.

\vspace{0.5cm}


\begin{Large}
 {\bf 4. Numerical analysis}
\end{Large}
 \vspace{0.5cm}

As mentioned in the preliminary survey of Section 2, among the potentially
relevant SUSY three-body modes (\ref{eq:processes}), only a few result in
sizeable corrections to the top quark width.
The relevant modes are defined to be as those that can give a contribution
\beq
{\delta\Gamma^{SUSY}\over \Gamma_{SM}}\stackM 1\%\,,
\label{eq:condition}
\eeq
with respect to the the canonical
width (\ref{eq:treeW}) of the top quark in the SM. 
For comparison, let us recall the size of the one-loop corrections to the
canonical decay (\ref{eq:canonical}) --these quantum corrections being of the
same order in perturbation
theory as our three-body decays at the tree-level. The corresponding SM electroweak  
corrections lie in the ballpark of
$1.5\%$ (in the $G_F$-scheme)\,\cite{TopSM}
for the present values of the top-quark mass, whereas the
QCD corrections are
of order $-(8-9)\%$\,\cite{TopSM} and are essentially independent of $m_t$.  
Our analysis shows that the SUSY three-body 
decays on eq.(\ref{eq:processes})
that most likely could fulfil condition (\ref{eq:condition}) 
are the modes I, IV, VIII, IX, X. 

As an independent set of SUSY inputs we introduce a similar tuple of 
parameters as in the numerical analyses of Refs. \cite{GJS2}-\cite{GS2}, specifically:
\beq
(\tan\beta, m_{H^+}, \mu, M, m_\sg, m_{\tilde{b}_1}, m_{\tilde{t}_1}, M_{LR}^b,
M_{LR}^t)\,.
\label{eq:inputsSUSY}
\eeq
For the full numerical analysis of the ten decays (\ref{eq:processes}) we have
produced several hundred plots in order to explore all the relevant peculiarities
of the $9$-dimensional parameter space of
eq. (\ref{eq:inputsSUSY})\,\cite{Jaume}.
In what follows we limit ourselves to report on 
the main results obtained under more simplified assumptions which do not alter 
the maximal rates expected. In particular, throughout all our numerical analysis
we use a representative set of inputs which is non-critical, i.e. such that our
results behave smoothly under variations around these values.
The simplifying assumptions mentioned above are the following.
We shall assume that the mixing angle
in the stop sector is $\theta_t=\pi/4$. In the sbottom sector, where mixing
effects are more unlikely, we fix
$M_{LR}^b=0$ and
$m_{\tilde{b}_1}=m_{\tilde{b}_2}\equiv m_{\tilde{b}}$.  
With these settings the stop mass eigenvalues $m_{\tilde{t}_{1,2}}$ are determined
and hence a more restricted set of parameters is obtained than the
general set (\ref{eq:inputsSUSY}). In most processes, this suffices.
Notwithstanding, we shall come back to the general set (\ref{eq:inputsSUSY}) later
on for some of the most promising decays.

Apart from the phenomenological limits on the various
sparticle masses mentioned above, we shall explore the crucial
parameter $\tan\beta$, eq.(\ref{eq:v2v1}), within the
range
\beq
0.5\stackm \tan\beta \stackm 70\,,
\label{eq:tanlimits}
\eeq
which is fixed by the perturbative domain of the Yukawa couplings
 (\ref{eq:Yukawas}).
Apart from the well established SM inputs\,\cite{PDB}, we adopt the following
for our numerical analysis:
\beq
\ba{c}
m_b =  5 \GeV \\
m_t =  175 \GeV \\
V_{tb} = 0.999 \\
\alpha_s(m_t) =  0.11\,.
\ea
\eeq
\begin{itemize}

\vspace{0.5cm}
\item{\bf Decay I}: There are two Feynman diagrams  contributing to this
process (Cf. Fig.1). Only two parameters from the general set (\ref{eq:inputsSUSY})
are needed in this case, namely
\beq
(\tan\beta, m_{H^+})\,.
\label{eq:plane}
\eeq 
Diagram (a) in Fig.1 is the conventional diagram, whereas diagram (b) gives the
extra Higgs contribution, which is determined by the strength of
the Yukawa couplings given on eq.(\ref{eq:Yukawas}) with
$m_{\tau}$ in place of $m_b$. Notice that in this case the charged Higgs need not
to be supersymmetric but just the charged member of a general two-Higgs doublet
model of type II\,\cite{Hunter}.
Since  $m_\tau> m_q$ for any light quark, in the region
of $\tan\beta>1$ we do not consider
the similar decays in which $(\nu_\tau,\tau^+)$ is replaced with $(u,\bar d)$
or  $(c,\bar s)$.
On imposing the condition that the two-body decay $t\rightarrow H^+\,b$ is not
kinematically permissible, we study the three-body decay only for 
Higgs masses 
\beq 
m_\hplus > m_t -m_b \,.
\label{eq:condit}
\eeq
Since we aim at a departure from the SM contribution, we define the quantity
\beq
\delta_I=\frac{\Gamma\left(t \rightarrow b \nu_\tau \tau^+\right) -
 \Gamma\left(t \rightarrow b W^{+ *} \rightarrow b \nu_\tau \tau^+\right)}
{\Gamma\left(t \rightarrow b W^{+ *} \rightarrow b \nu_\tau \tau^+\right)}\,,
\label{eq:delta}
\eeq
and plot contour lines of $\delta_I$ in the plane (\ref{eq:plane})  under
the condition (\ref{eq:condit}). The result is displayed in Fig. 9. Notice that 
$\Gamma\left(t \rightarrow b \nu_\tau \tau^+\right)$ on eq.(\ref{eq:delta}) is
computed from the sum of the two amplitudes in Fig.1, whereas 
$\Gamma\left(t \rightarrow b W^{+ *} \rightarrow b \nu_\tau \tau^+\right)$
includes the first amplitude only. It is seen from Fig. 9 that corrections of a
few percent are possible at high $\tan\beta$.
In the low $\tan\beta<1$ region, the decay under study is inefficient since the
$\tau$-lepton Yukawa coupling becomes very depleted.
In spite of the fact that in this region of $\tan\beta$ the alternative decay
$t \rightarrow b\,c \bar{s}$ can give contributions of order $1\%$, this
channel would be much more difficult to separate from the background. 

The dominance of the
the semileptonic channel for $\tan\beta>{\cal O}(1)$ occurs for both real or virtual Higgs
decays. For real decays, we have
\beqn  
{\Gamma(H^{+}\rightarrow\tau^{+}\nu_{\tau})\over
\Gamma(H^{+}\rightarrow c\bar{s})}&=&\frac{1}{3}
\left(\frac{m_{\tau}}{m_c}\right)^2
{\tan^2\beta\over
(m^2_s/ m^2_c)^2\tan^2\beta+\cot^2\beta}\nonumber\\
&\rightarrow & \frac{1}{3}\left(\frac{m_{\tau}}{m_s}\right)^2
>10\ \ \ \ \ \ ({\rm for}\ \tan\beta>\sqrt{m_c/m_s}\stackM 3)\,.
\label{ratiotaucs}
\eeqn
The identification of the $\tau$ mode 
could be a matter of measuring a departure from the universality prediction
between all lepton channels.  
Fortunately, $\tau$-identification is possible at
Tevatron\,\cite{tauTev}; and the feasibility of tagging the excess of events
with one isolated $\tau$-lepton as compared to events with an additional lepton
has been substantiated by studies of the LHC collaborations \,\cite{Atlas}.
It has recently been shown that it should be fairly easy to  
discriminate between the $W$-daughter $\tau$'s and the 
$H^{\pm}$-daughter $\tau$'s by just looking at
the opposite states of $\tau$ polarization resulting from the $W^{\pm}$
and $H^{\pm}$ decays; the two polarization states can be distinguished by
measuring the charged and neutral contributions to the $1$-prong $\tau$-jet
energy (even without identifying the individual meson states)
\,\cite{Raychau}.

\vspace{0.5cm}
\item{\bf Decay II}: There are four Feynman diagrams contributing to this decay,
and they are displayed in Fig.2. In principle, $h^0$ in these diagrams could be 
replaced with any
of the neutral MSSM higgses. However, 
the following relations must hold simultaneously
\beq
\ba{lcccl}
m_{h^0}& <& m_t-m_b-M_W\\
m_\hplus &>& m_t-m_b\,.
\ea
\label{eq:2rel}
\eeq
in order to allow the three-body decay and at the same time to
forbid the two-body charged Higgs mode (\ref{eq:tHb}) involved
in one of the amplitudes.
In the MSSM, the second relation implies (at the tree-level)\,\cite{Hunter}
that $m_{A^0}>155\,\GeV$,
and since the heavy CP-even Higgs satisfies $m_{H^0}>M_Z$, the relations 
(\ref{eq:2rel})
can only be fulfilled by the lightest CP-even Higgs $h^0$.  Again the only
two parameters involved are those in eq.(\ref{eq:plane}), since the 
mixing angle, $\alpha$, between the CP-even higgses is not independent 
in the MSSM\,\cite{Hunter}. Unfortunately, a fully fledged calculation of the
diagrams in Fig.2 yields a disappointingly small contribution, as seen in Fig.10.
The reason for it is the following.
The potentially relevant neutral Higgs Yukawa couplings  $h^0\,t\,t$ 
and $h^0\,b\,b$ in Figs. 2a and 2c are proportional to
$\cos\alpha/\sin\beta$ and $\sin\alpha/\cos\beta$, respectively\,\cite{Hunter}. 
Since $m_{A^0}$ is assumed to
be large, then $\alpha$ goes to zero or to $-\pi/2$ depending on whether
$\tan\beta$ is large or small, respectively. In neither case these
Yukawa couplings become enhanced with respect to the gauge coupling.
The other potentially
relevant interaction is the charged Higgs vertex $H^+\,t\,b$ in Fig. 2b, 
which is sensitive to both Yukawa couplings on eq.(\ref{eq:Yukawas}) and is
independent of $\alpha$\,\cite{Hunter}. Unfortunately, this vertex
is also rendered ineffective
in the present conditions, since its behaviour at large and small $\tan\beta$
is compensated for by the opposite behaviour of the companion 
vertex $W^+\,\,H^+\,h^0$ in the same diagram, which
is proportional to $\cos (\alpha-\beta)\rightarrow 0$.
In the end there is a balance between the four diagrams giving rise to
a maximum near $\tan\beta=1.5$, although the resulting contribution
with respect to the canonical decay is  rather meagre, namely
\beq
\delta_{II}=\frac{\Gamma(t \rightarrow h^0\, b\, W^+)}
{\Gamma_{SM}} < 3\times 10^{-4}\,,
\eeq
making the observation of this process essentially hopeless.  

\vspace{0.5cm}
\item{\bf Decay III}: The corresponding Feynman diagrams are in
 Fig.3\,\footnote{We remark that in all our Feynman diagrams indices 
for virtual sparticles are understood to be summed over.}.
Obviously, this decay can only proceed within the context of the so-called 
light gluino scenario
$m_{\tilde{g}}={\cal O}(1)\,\GeV$\,\cite{Clavelli}, since for heavy gluinos the
usual bound (\ref{eq:hgluino}) prevents the decay from occurring. 
With $m_{\tilde{g}}$ assumed to be negligible, we are left
with the additional conditions
\beq
\ba{c}
m_\sbt< m_t-M_W\\
 m_\stp> m_t\,,
\ea
\eeq
where the second relation is to guarantee that the two-body decay (\ref{eq:tt1g})
does not take place. The relevant subset of parameters in the present instance is
\beq
(\tan\beta, m_\sg,  m_\sbt, M_{LR}^t)\,.
\eeq
The numerical analysis is presented for a light gluino
mass $m_{\tilde{g}}=1\,\GeV$ and fixed $\tan\beta=1$ since
small values of $\tan\beta$ are preferred. The results are
basically the same within the light gluino range $m_{\tilde{g}}=(1-10)\,\GeV$  
and are rather insensitive to allowed values of $M_{LR}^t$  
below $m_t$, so we have set $M_{LR}^t=0$.
In Fig. 11 we display the contribution of the LH sbottom final state. 
For LH sbottom masses respecting the
absolute LEP bound (\ref{eq:LEP140}), we find                                                                                                                                                    
\beq
\delta_{III}=\frac{\Gamma(t \rightarrow \tilde{b}_a
\, W^+\, \sg)}{\Gamma_{SM}} < 0.1\,,
\eeq
which is well below the limit (\ref{eq:condition}) and therefore is too small.
The yield from
the RH sbottom final state is even smaller ($<10^{-5}$) due to the helicity
flip at the bottom line in Fig. 3a.

\vspace{0.5cm}
\item{\bf Decay IV}:This process is rather complex since it involves 
four Feynman diagrams (see Fig. 4) and the following parameters:
\beq
(\tan\beta, m_{H^+}, \mu, M, m_{\tilde b}, M_{LR}^t)\,.
\label{eq:inputsIV}
\eeq
The simultaneous conditions to be required in order that this decay is possible
while the two-body modes (\ref{eq:tb1Psi1})-(\ref{eq:tHb}) are forbidden are
the following:
\beq
\ba{c}
M_{\chi^0_{\alpha}} + M_{\chi^+_i} + m_b < m_t\\
m_\hplus > m_t-m_b\\
M_\neut + m_\st >m_t\\
M_\cplus + m_\sbt > m_t\,.\\
\ea
\label{eq:relsIV}
\eeq
Notice that in practice the handling of these relations has to be 
carried out numerically since
our inputs (\ref{eq:inputsIV}) are not given directly in terms of the chargino and
neutralino masses. Therefore one has to diagonalize the chargino and neutralino
mass matrices on eq.(\ref{eq:CNMM}) and look for regions in the higgsino-gaugino
parameter space $(\mu, M)$ where the relations (\ref{eq:relsIV}) are met.
We have checked that the maximum contribution of this decay is obtained
near the phenomenological boundaries
defined by the condition (\ref{eq:LEP140}).
In Fig. 12 we plot the ratio 
\beq
\delta_{IV}={\Gamma(t \rightarrow b\, \chi^0_1\,
\chi^+_1)\over \Gamma_{SM}}\,,
\label{eq:contour}
\eeq
as a function of $\tan\beta$ for 
a typical $(\mu, M)$ boundary point  
corresponding to $m_{\cplus}=60\,\GeV$.
We see that
at high $\tan\beta$ there are significant enhancements of $\delta_{IV}$
which could boost this quantity up to a value $\sim 6-8\%$, i.e. up to the level of 
the conventional QCD corrections to the canonical decay.
We have checked that
the total effect from the next-to-lightest charginos and
neutralinos allowed by phase space
amounts to an additional contribution of $0.6\%$ in the high $\tan\beta$ region.   
In view of these results, it follows that this
three-body mode could be relevant.

\vspace{0.5cm}
\item{\bf Decay V}: Three Feynman diagrams contribute to this decay (see Fig. 5)
and the parameters involved read
\beq
(\tan\beta, \mu, M, m_\sg, m_{\tilde b}, M_{LR}^t)\,,
\label{eq:inputsV}
\eeq
which are limited by the conditions
\beq
\ba{c} 
m_\st < m_t-m_b- m_\sbt\\ 
m_\sg > m_t-m_\st \\ 
M_\neut+m_\st > m_t\\
M_\cplus + m_\sbt > m_t\,.
\ea
\eeq
Since gluinos can either be very light, viz. of ${\cal O}(1)\,\GeV$, or 
at least as heavy as
$100\,\GeV$, the second condition above enforces a heavy gluino scenario.
Within our current set of hypotheses we exclude a heavy stop and a light gluino
since the (degenerate) sbottoms would be heavy, too, and the phase space
would be exhausted. In Fig. 13 we present a ``lattice plot'' sampling\,\cite{GJS2} 
of the $6$-tuple (\ref{eq:inputsV}), 
separately for LH and RH sbottoms.  We see that the ratio
\beq
\delta_{V}={\Gamma(t \rightarrow \stp\, \tilde{b}_a\, \bar b)\over \Gamma_{SM}}\,,
\eeq
shows the highest cumulative number of points for values $\delta_{V}<0.2\%$
whereas higher values get only a few number of spots. 

\vspace{0.5cm}
\item{\bf Decay VI}: This decay is particularly simple, for it admits a single
Feynman diagram (Fig. 6a). The input parameters are
\beq
(\tan\beta, \mu, M, m_{\tilde{b}}, m_{\tilde s})\,,
\label{eq:inputsVI}
\eeq
under the conditions
\beq
\ba{c} 
2\,m_\sbt< m_t-m_c \\
m_\sbt+M_\cplus > m_t\,,
\ea
\eeq
where we assume $m_{\tilde{b}}=m_{\tilde{s}}$. Thus we are led to an scenario
of heavy charginos and relatively light squarks. In Fig. 14a we display the ratio
\beq 
\delta_{VI}={\Gamma(t \rightarrow \tilde{b}_a\, c\, 
\bar{\tilde{s}}_b)\over \Gamma_{SM}}\,,
\eeq 
for the different combinations of chiral species of squarks. Contributions
of order $\delta_{VI}\simeq 1\%$ obtain for $\tan\beta\stackm 0.6$, 
i.e. near the lower limit (\ref{eq:tanlimits}).

\vspace{0.5cm}
\item{\bf Decay VII}: This decay is similar to the previous one  (see Fig. 6b), 
but it involves $\tau$-sleptons in the final state. Thus the inputs are
\beq
(\tan\beta, \mu, M, m_{\tilde{b}}, m_{\tilde\tau})\,,
\label{eq:inputsVII}
\eeq
bounded as follows
\beq
\ba{c} 
m_\sbt+m_{\tilde\tau}< m_t \\
m_\sbt+M_\cplus > m_t\,.
\ea
\eeq
The relevant ratio
\beq 
\delta_{VII}={\Gamma(t \rightarrow \tilde{b}_a\, \nu_{\tau}\,
 \tilde{\tau}_b^+)
\over \Gamma_{SM}}\,,
\eeq
is analyzed in Fig. 14b
for the different combinations of chiral species of staus and sbottoms.
In the present instance, where sleptons are around,  we assume
a corresponding mass matrix with the same simplified
structure as that of the sbottom mass matrix. Therefore, the LH and RH staus
are degenerate, with a common mass 
$m_{\tilde{\tau}}\stackM 95\,\GeV$, which guarantees
$m_{\tilde{\nu}}\stackM 40\,\GeV$ -- as required 
by the data on the invisible $Z$-width\,\cite{LEPEWWG}. In these 
conditions, it is seen from Fig. 14b that $\delta_{VII}\simeq 1\%$ is achieved 
for $\tan\beta$ very near the two extreme values of the allowed interval
(\ref{eq:tanlimits}), but only
at the expense of a rather light chargino $M_{\cplus}\simeq 50\,\GeV$,
which is still permitted if it is almost degenerate with the lightest
neutralino\,\cite{LEP140}.

\vspace{0.5cm}
\item{\bf Decay VIII}: This decay is related to number IV in that a
gluino substitutes for a neutralino. The number 
of Feynman amplitudes stays the same (Cf. Fig. 7) and the relevant tuple of
independent parameters can be chosen as follows:
\beq
(\tan\beta, m_\sg, \mu, M, m_{\tilde b}, M_{LR}^t)\,,
\label{eq:inputsVIII}
\eeq
being subdued by the conditions
\beq
\ba{c}
M_{\chi^+_i}+m_\sg < m_t-m_b\\
m_\st + m_\sg > m_t \\
m_\sbt + M_{\chi^+_i} > m_t\,,
\ea
\eeq
whose meaning should by now be pretty obvious. In contrast to decay V,
both the light and heavy gluino scenario is permissible in the present instance.
In the first case (light gluinos), a heavy stop and a heavy chargino could coexist
with a relatively light sbottom, while in the second case (heavy gluinos) a light stop
is tolerated at the expense of heavy sbottoms. The numerical analysis of the
quantity  
\beq
\delta_{VIII}={\Gamma (t \rightarrow b\, \sg\, \chi^+_i)\over\Gamma_{SM}}\,,
\eeq
shows that in the 
heavy gluino scenario the contributions are well
below $1\%$ and we do not show explicit plots.
In contrast, for light gluinos we see in Fig.15 that  $\delta_{VIII}$
could border on values of order $1\%$ for any value of $\tan\beta$, and 
it could even reach $4\%$ for sufficiently 
small or big  values of $\tan\beta$ within the interval (\ref{eq:tanlimits}).
This result is
quite remarkable, since it is about $60\%$ of the QCD corrections to the
canonical decay. 

\vspace{0.5cm}
\item{\bf Decay IX}: This is a very interesting three-body decay to
deal with, though its analysis is technically quite demanding for it involves
four Feynman amplitudes (see Fig. 8). 
In the Appendix at the end of the paper we present the full squared amplitude 
corresponding to this particular decay.
The parameter space includes as many as nine parameters, namely
\beq
(\tan\beta, m_{H^+}, \mu, M, m_\sg, m_{\tilde{b}_{1,2}},
m_{\tilde{t}_1}, M_{LR}^t)\,.
\label{eq:inputsIX}
\eeq
In this case, and due to the potentiality of this mode, we relax the
assumption on the stop mixing angle being $\theta_t=\pi/4$ and
introduce the lightest stop mass $m_{\tilde{t}_1}$ as a new input.
Furthermore, we also abandon the restriction $M_{LR}^b=0$ and the assumption
of degenerate sbottom masses by making allowance for a free input value of
the sbottom mass eigenvalues $m_{\tilde{b}_{1,2}}$
with the proviso that $\theta_b=\pi/4$. 
In contrast to the previous decays, we shall not impose conditions 
blocking the possibility that the intermediate two-body states in Fig.8 
can be real two-body decays, except for the gluino mode
$t\rightarrow \tilde{t}\,\tilde{g}$ which, if allowed, would be overwhelming
in most of the parameter space. 
As for the remaining two-body modes, it turns out that the present
three-body decay can be competitive with them in certain regions of parameter
space.

The numerical analysis of decay IX confirms the 
expected fact that it can be relevant only if the two-body channel 
$H^+\rightarrow\tilde{t}\,\bar{\tilde{b}}$ is kinematically forbidden, otherwise the
Higgs decay width becomes too large and it has a dramatic suppression effect on
the partial width of our three-body decay. This can be seen in Fig.16a, corresponding
to $m_\st=60\,\GeV$, $m_\sbt=100\,\GeV$, $m_\sg=130\,\GeV$ and a large value of 
$\tan\beta$ of order $m_t/m_b\simeq 36$. We have also fixed $\mu=M=100,250\,\GeV$
and $M_{LR}^b=0$.
We see indeed that, for $m_{H^+}$ approaching $m_\st+m_\sbt$ from below, the decay IX
can give a contribution to the ratio  
\beq
\delta_{IX}={\Gamma (t\rightarrow b\, \tilde{t}_1\, \bar{\tilde{b}}_1)
\over\Gamma_{SM}}\,,
\eeq
which ranges from a few percent to $100\%$ and above.
The contribution from $\tilde{b}_2$ is basically
the same.
Clearly, this behaviour is unmatched so far by
any of the previous decays. 
Therefore, these non-standard effects are  
to be seriously considered in any consistent treatment of the total top quark
width. We postpone for Section 5 the discussion of the
possible signatures.

Since in the conditions under study, the two-body mode $t\rightarrow H^+\,b$ is still
available, we also plot in Fig.16a the ratio
\beq
{\Gamma (t\rightarrow H^+\,b)\over\Gamma_{SM}}\,.
\eeq
In this way we see that, depending on the value of $\mu$, there may be a
sizeable range of Higgs masses where the decay IX is not negligible as 
compared to the decay  $t\rightarrow H^+\,b$. 
Although these results have been obtained by assuming that the two sbottoms are
degenerate, we may also use the general set of inputs
(\ref{eq:inputsIX}) to enter different values
for the sbottom masses. For example, take $m_{\tilde{b}_1}=100\,\GeV$ and
$m_{\tilde{b}_2}=150\,\GeV$, and set $\mu=100\,\GeV$
without altering the rest; we then obtain a rate 
$\delta_{IX}=33\%$. In all these cases the soft SUSY-breaking parameter
$A_b$ is necessarily large to avoid 
conflict with the phenomenological mass bounds. In fact,
if in the previous example we would trade 
the input parameter $\mu$ 
for $A_b$, we would find that we cannot demand a small input value 
$A_b\simeq 0$ since it would entail a too small a value of
$\mu$ below $50\,\GeV$, which is ruled out
by the chargino mass bound (\ref{eq:LEP140}).

Since the general set (\ref{eq:inputsIX}) is found not to change the order
of magnitude
of the rate of our decay, it will simplify the analysis to restore the
original set of inputs
\beq
(\tan\beta, m_{H^+}, \mu, M, m_\sg, m_{\tilde{b}}, M_{LR}^t)\,.
\label{eq:inputsIX2}
\eeq
where $\theta_t=\pi/4$, $M_{LR}^b=0$ together with the assumption of
degenerate sbottom masses.   
In Figs.16b and 16c we study the evolution of $\delta_{IX}$ versus $\tan\beta$ 
for the two regimes where the Higgs decay
 $H^+\rightarrow\tilde{t}\,\bar{\tilde{b}}$
is closed and open, respectively. 
In the closed case the slope is
very high in the large  $\tan\beta$ region. As a consequence,  
the yield from the Higgs mediated diagram 
becomes so overwhelming at large $\tan\beta$ that
the rate of the decay IX overtakes that of the canonical
decay as soon as $\tan\beta\stackM 50$,!.
Notice that the long flat region in the intermediate interval
$1\stackm\tan\beta\stackm 20$ is sustained by the 
gluino mediated diagram in Fig.8, though in this plateau the contribution
stays small due to the large mass assumed for the gluino, which prevents the
$t\rightarrow\tilde{t}\,\tilde{g}$ decay from occurring.
Whereas the dependence on the soft SUSY-breaking parameter $M$ is very
mild, in Fig. 16d we
exhibit the dramatic dependence on the parameter $\mu$ for typical
values of the other parameters. As already mentioned, the region
 $\mu\stackm 60\,\GeV$ (which is unfavoured by our decay)
is excluded by the present bounds on chargino masses.
Notice also that when the real Higgs decay into $\tilde{t}\,\bar{\tilde{b}}$
is viable
the ratio $\delta_{IX}$ lessens considerably, viz. 
down a few percent level and mostly at a few per mil or below. 

Intriguingly enough, there are regions of parameter space where the decay
IX can be
competitive with the main two-body modes available.
In Figs. 17a and 17b we plot the two-body  decay rates 
\beq
{\Gamma (t\rightarrow\tilde{t}_1\,\chi_{\alpha}^0)\over\Gamma_{SM}}\,,\ \ \ \ \
{\Gamma (t\rightarrow\tilde{b}_a\cplus)\over\Gamma_{SM}}\,,
\label{eq:twobody}
\eeq
as a function of $\tan\beta$
for two values of $\mu$ and for fixed $M_\cplus=60\,\GeV$. 
For the squark masses we have taken $m_{\tilde{b}_a}=100\,\GeV$,
$m_{\tilde{t}_1}=60\,\GeV$.
In the first decay we have
included separated plots for the first three neutralinos $\chi_1^0, \chi_2^0,
\chi_3^0$, since they are also allowed by
phase space. Although the last two are heavier than $\chi_1^0$ 
they are worth including, the reason being that
they are more higgsino-like and therefore have larger Yukawa couplings of the form
(\ref{eq:Yukawas}) resulting in a competitive contribution.
For the second decay in (\ref{eq:twobody}), the chargino $\chi_2^+$ has not been
included since it is too heavy.
It is seen that the
rates can be high for small $\tan\beta$, and in this region they are more efficient 
than $\delta_{IX}$. However, for $\tan\beta> 30$,
the three-body decay 
width $\delta_{IX}$
is not only competitive but it can even surpass the rate of the previous two-body
decays, especially for large enough $\mu$ where the partial width of the
former increases substantially whereas the two-body partial widths
(\ref{eq:twobody}) decrease.

\vspace{0.5cm}
\item{\bf Decay X}: This decay is similar to the previous one. The Feynman
diagrams are similar to those in Fig. 8 by just replacing 
$\tilde{t}\rightarrow\tilde{\nu}_{\tau}$ and $\tilde{b}\rightarrow\tilde{\tau}$
and forgetting about gluino and neutralino mediated amplitudes.
The parameter space for this decay reads
\beq
(\tan\beta, m_{H^+}, \mu, M_{LR}^{\tau}, m_{\tilde{\nu}_{\tau}}).
\label{eq:inputsX}
\eeq
An obvious restriction to be fulfilled is
\beq
m_\sneut < m_t-m_b- m_{\tilde{\tau}}\,.  
\eeq
For simplicity we shall assume that the two $\tilde{\tau}$-sleptons
are degenerate in mass and that 
$ M_{LR}^{\tau}=0$. As in the decay IX, we expect that the ratio 
\beq
\delta_{X}={\Gamma (t\rightarrow b \sneut_\tau \tilde{\tau}^+_1)\over\Gamma_{SM}}\,,
\eeq
will be most sizeable if the two-body Higgs decays
$H^+\rightarrow \sneut_\tau \tilde{\tau}^+$ and
$H^+\rightarrow\tilde{t}\,\bar{\tilde{b}}$
are kinematically forbidden. 
If this is so, for $m_{H^+}$ approaching $m_{\tilde{\nu}_{\tau}}+m_{\tilde{\tau}}$ 
from below, the decay X can furnish a contribution to the ratio $\delta_{X}$
which ranges
between the few percent to the several ten percent, depending on the value of the
other parameters, specially of $\tan\beta$ and $\mu$. This is seen in
Fig.18a, where the relevant inputs are 
\beq
(\tan\beta, \mu, m_{\tilde{\nu}_{\tau}})=(40, 100\,\GeV, 50\,\GeV).
\eeq
Cases $\tilde{\tau}_1$ and $\tilde{\tau}_2$ are almost indistinguishable
since $\tan\beta$ is high in Fig. 18a.
In Figs.18b and 18c we plot the evolution of $\delta_{X}$ on $\tan\beta$ and on
$\mu$, respectively, for fixed values of the other parameters.
 Notice that on increasing $\tan\beta$
the Higgs coupling to stau and the corresponding sneutrino increases but at the same
time the stau mass also increases. For $m_{\tilde{\nu}_{\tau}}=50\,\GeV$, this mass
saturates at a value of $m_{\tilde{\tau}}\simeq 92\,\GeV$ for $\tan\beta\simeq 10$.
Therefore, for larger values of $\tan\beta$, the ratio $\delta_{X}$ steadily grows 
and it can surpass $50\%$ for $\tan\beta\stackM 55$. 
These results can still be improved substantially for higher values
of $|\mu|$ (Cf. Fig.18c).

\end{itemize}

\vspace{0.5cm}


\begin{Large}
 {\bf 5. Discussion and conclusions}
\end{Large}
 \vspace{0.5cm}

To summarize, we see that there are some three body decays of the top quark within
the MSSM whose contribution can be a fraction of the standard width
$\Gamma_{SM}=\Gamma(t\rightarrow W^+\,b)$
of order or above $1\%$. The latter reference value is approximately the size of the
conventional electroweak
quantum corrections to $\Gamma_{SM}$\,\cite{TopSM}.
For the sake of comparison, let us point out that the Higgs effects on 
$t\rightarrow W^+\,b$ within the MSSM are about one order of magnitude smaller
than our reference value, i.e. they are of order $0.1\%$\cite{GHDH}. 
Many of our three-body decays could give a contribution above this small number.
In general, however, we find that only a few number of these three-body decays  
do contribute, within the same order of perturbation theory, as much as
the full (electroweak plus QCD) quantum effects on $t\rightarrow W^+\,b$
in the SM, which are of order $-7\%$\,\cite{TopSM}.
Nonetheless, if we consistently add up the yield from some of these decays,
we are able to find regions of the parameter space where the resultant
pay-off could counterbalance the one-loop quantum effects. This could  be
the case e.g. if we add the contributions from decays I and IV within the
heavy gluino scenario, or decays I and VIII within the light gluino scenario.
In either of these situations the one-loop corrections would appear as
``missing effects'' in a measurement of the total width of the top quark,
$\Gamma_t$. 
This feature is remarkable, since it is compatible with an scenario in which all
of the SUSY two-body decays of the top quark are blocked up. 
In such a case the leading SUSY signature in {\it inclusive} top physics could come
from the combined effect of the available three-body decays.     

Finally, there are two cases, viz. decay IX and decay X, which could 
be very important in certain regions of parameter space, specifically in a region
where both $\tan\beta$ and $|\mu|$ are large enough,
$\tan\beta\stackM m_t/m_b\simeq 36$ and $|\mu|>100\,\GeV$, provided that the
Higgs mass is below the  
$\sneut_\tau \tilde{\tau}^+$ and $\tilde{t}\,\bar{\tilde{b}}$ thresholds.
In this case,
the last two decays could contribute to the total top quark width a fraction
which, if one does not stretch too much the values of the
relevant parameters,  it can still
range between the few percent to the several ten percent.
The large size of these contributions are the tree-level counterpart, within the
same order of perturbation theory, of the 
large one-loop quantum effects
on the two-body decay (\ref{eq:tHb})\,\cite{GJS,CGGJS}\footnote{Notice 
that some of the Yukawa couplings (\ref{eq:Yukawas}) and triple scalar couplings
(\ref{eq:gij}) can be comparable to the strong gauge coupling (\ref{eq:Lqsqglui}).
Therefore, the tree-level corrections to decay IX can be of
the same order as the SUSY-QCD\,\cite{GJS} and electroweak SUSY 
corrections\,\cite{CGGJS} to the two-body decay (\ref{eq:tHb}).}.
Therefore, decays IX and X  
could possibly be searched as exclusive decays since they may
give rise to well defined and rather exotic signatures.
These signatures have to be compared with the canonical signature
\be
t\rightarrow b W^+ \rightarrow b (f \bar f')\,,
\label{eq:decstan}
\ee
which consists either of 3~jets (one of them a b-jet)
or a ~b-jet~+ a positively charged lepton (~$l^+$~)+ missing 
energy-momentum (~$\slas{p}$).
In Table 1 we display the
alternative decay signatures associated to decay IX.
In all cases we have obviated
an additional b-jet which is in common with the canonical signature.
Therefore, what we
show is the specific signature associated to the
virtual decay $H^+\rightarrow\tilde{t}\,\bar{\tilde{b}}$\,\cite{Bartl}. 
In some of the signatures in Table 1 (cases (a)-(c) )
we have admitted of the exotic possibility to having FCNC decays 
mediated by neutralinos
\,\cite{Cuypers}.
These could enter the game if the 
chargino decay mode of the stop,
\beq
\st_1 \rightarrow b \cplus\,,
\label{eq:stopcharg}
\eeq
is kinematically
forbidden. If this mode is allowed, it would be dominant and the main signatures
would be (d)-(f) in Table 1. As for the sbottom decay, the charged mode
$\bar\sbt \rightarrow \bar c \cplus$ is CKM-suppressed whereas
$\bar\sbt \rightarrow \bar b \neut$ is always available (if a light neutralino
exists). Let us point out that if the
former case applies it would lead to two final leptons with the
same electric charge (see case (f)), which is definitely a non-canonical signature.  
In general,  we should expect that the leading signatures in Table 1
are (a)-(b) or (d)-(e),
depending on whether the channel (\ref{eq:stopcharg}) is closed 
or open, respectively.

As far as decay X is concerned, a similar discussion ensues. The main signatures
are detailed in Table 2. It could well happen that the charged sneutrino decay
\beq
\sneut_\tau \rightarrow \tau^- \cplus\,,
\label{eq:sneutcharg}
\eeq
is phase space obstructed, similarly to the previous case (\ref{eq:stopcharg})
for the stop. However, an important difference here is that the neutral decay
\beq
\sneut_\tau \rightarrow \nu_\tau \neut\,,    
\label{eq:sneutneut}
\eeq
is perfectly possible and most likely it is the leading decay mode of the
sneutrino, if $\neut$ is the lightest SUSY particle.
Another difference with respect to decay IX is that whereas the
charged decay of the sbottom is CKM-suppressed, the charged
$\tilde{\tau}$-decay $\stau^+ \rightarrow \bar\nu_\tau \cplus$ is not.
We conclude that the leading signatures are (a)-(c) in
Table 2, where the characteristic trait is the proliferation of
jets, lepton prongs
and missing
energy. However, if the charged decay mode (\ref{eq:sneutcharg}) 
would be available, it
could be rather interesting, for the corresponding signatures (d)-(f) in Table 2
show the distinctive presence of a $\tau$-lepton with the ``wrong'' sign, i.e  a
$\tau^-$, in contradistinction to the canonical 
decay (\ref{eq:decstan}) where one expects to find a $\tau^+$. Notice that 
in the conditions under study the signatures on Tables 1 and 2 cannot come
from the decay of a real $H^+$; in fact, although real charged Higgs
bosons can be produced in
these conditions, their leading decay would be
$H^+\rightarrow\tau^+\,\nu_{\tau}$. 
With enough statistics, the latter signature would lead to a violation
of lepton universality and it has a chance to be isolated in the 
future (Cf. Section 4); however,
with an insufficient statistics it could easily be confused with a 
standard signature (\ref{eq:decstan}).
Quite in contrast, the exotic signatures displayed in Tables 1 and 2 are
highly non-standard and may play an important role in the top quark 
phenomenology beyond the SM, especially in the analysis of
the single top-quark production
processes\,\cite{Heinson}, which are foreseen to play a decisive role
in the near future at Tevatron and at the LHC.


{\bf Acknowledgements}:

\noindent
One of us (J.S.) thanks A. Heinson for providing useful information on
top quark physics at the Tevatron.
This work has been partially supported by CICYT 
under project No. AEN95-0882. The work of J.G. has also been financed by a
grant of the Comissionat per a Universitats i Recerca, Generalitat de
Catalunya.

\vspace{0.75cm}

\begin{Large}
{\bf Appendix A}
\end{Large}
 \vspace{0.5cm}
\setcounter{equation}{0}
\renewcommand{\theequation}{A.\arabic{equation}}
 
In this appendix we write out the full analytical expression for the squared
amplitude of one single decay, since the complete formulae for all the
processes that have been analyzed is extremely lengthy. We have chosen
the decay IX, which is one of
the most relevant candidates in the list of SUSY three-body modes presented
on eq.(\ref{eq:processes}).  
Using the Lagrangian interactions and coupling matrices defined on eqs.
(\ref{eq:Lqsqcn})-(\ref{eq:AS}), (\ref{eq:Lqsqglui})-(\ref{eq:gij}),
(\ref{eq:Wq})-(\ref{eq:WCN}),
along with the well-known SUSY Higgs-fermion interactions\,\cite{Hunter}, one 
computes the Lorentz invariant amplitudes $T_i\,(i=1,...,4)$ of
the decay $t\rightarrow b\, \st_b \,\bar\sbt_a\,\,(a,b=1,2)$ 
in terms of the four-momenta specified in Fig. 8a. 
The total squared amplitude summed over all final spin, colour and
sflavour indices and averaged over the initial states can be casted in the 
following way:
\beq
\sum \langle \left| T\right|^2 \rangle=
\sum_{i} Q_{ii}+\sum_{i < j} \left(Q_{ij}+Q_{ij}^\dagger\right)
\eeq
where   
\beq
Q_{ij}=\sum \langle T_i T_j^\dagger \rangle\,. 
\eeq
The explicit structure of the various $Q_{ij}$ is rather cumbersome. We first
do the following definitions\footnote{Here, for convenience, we express
our results in real, instead of complex, matrix notation following the
rules stated in Section 3.}:
$$ 
\ba{lcl}
\epsilon_W&=&M_W \Gamma_W\\
\epsilon_H&=&m_{\hplus} \Gamma_{\hplus}\\
G_{ab}&=&g_{cd} \,R^{(b)}_{da}\,R^{(t)}_{cb}=g_{LL}\,R^{(b)}_{1a}
\,R^{(t)}_{1b}+
g_{RR}\,R^{(b)}_{2a}\,R^{(t)}_{2b}+g_{LR}\,R^{(b)}_{2a}\,R^{(t)}_{1b}+
g_{RL}\,R^{(b)}_{1a}\,R^{(t)}_{2b}\\
Sum&=&\sum_{a,i,j,s,t} \lambda_{ij}^a \lambda_{st}^a=12
\ea
$$
and derive the following results:
\vspace{1.25cm}
$$
\begin{array}{lcl}
  Q_{aa}&=&{{3\,{g^4}} \,{{R^{(b)}_{1a}}^2}\, {{R^{(t)}_{1b}}^2}\over {4\,\left(
     {{ {{\left( {{{ q_1}}^2} -{{{ m_W}}^2} \right) }^2+{ \epsilon_W}}^2} }
     \right) }}\,\left( 2\,\left( { p_1k} - { p_2k} \right) \, \left( { p_1p_3}
     -{ p_2p_3} \right) \phantom{+ {{{{\left( -{{{ m_\sbt}}^2}\right) }^2}
     }\over {{{{ m_W}}^4}}} \left( {{{ q_1}}^2} \right) }\right.\\&&- \left( {{{
     m_\sbt}}^2} + {{{ m_\st}}^2} - 2\,{ p_1p_2} \right) \, { p_3k} \\&&-
     {{2\,\left( {{{ m_\st}}^2}-{{{ m_\sbt}}^2} \right) }\over {{{{ m_W}}^2}}}\,
     \left( { kq_1}\,\left( { p_1p_3} - { p_2p_3} \right) - \left( {{{
     m_\st}}^2} -{{{ m_\sbt}}^2} \right) \,{ p_3k} + \left( { p_1k} - { p_2k}
     \right) \,{ p_3q_1} \right) \\&&\left.+ {{{{\left( {{{ m_\st}}^2} -{{{
     m_\sbt}}^2} \right) }^2} }\over {{{{ m_W}}^4}}} \left( 2\,{ kq_1}\,{
     p_3q_1} - { p_3k}\,{{{ q_1}}^2} \right) \right) \,
\end{array}
$$
\vspace{1.25cm}
$$
\begin{array}{lcl}
  Q_{bb}&=&{{3\,{g^4}}\,{{ G_{ab} }^2}\over {4\,{{{ m_W}}^4}\, \left( {{\left(
   {{{ q_1}}^2} -{{{ m_\hplus}}^2} \right) }^2} + {{{ \epsilon_\hplus}}^2}
   \right) }} \, \\&& \left( 2\,{{{ m_b}}^2}\,{{{ m_t}}^2} + { p_3k}\,\left({{{
   m_t}}^2}\, {{{ \cot\beta}}^2}+ {{{ m_b}}^2}\,{{{ \tan\beta}}^2} \right)
   \right) \,
\end{array}
$$
\vspace{1.25cm}
$$
\begin{array}{lcl}
Q_{cc}&=&{{8\,{{{ g_s}}^4} } \over {3\,{{\left( {{{ q_2}}^2} -{{{ m_\sg}}^2}
    \right) }^2}}} \,\left( 2\,{ m_b}\,{{{ m_\sg}}^2}\,{ m_t}\,
    R^{(b)}_{1a}\,R^{(b)}_{2a}\,R^{(t)}_{1b}\, R^{(t)}_{2b} \phantom{ \left({
    p_3k} \right) \, \left( {{R^{(t)}_{2b}}^2} \right) }\right.\\&& + 2\,{
    m_b}\,{ m_t}\,{{{ q_2}}^2}\, R^{(b)}_{1a}\,R^{(b)}_{2a}\,R^{(t)}_{1b}\,
    R^{(t)}_{2b} \\&&- 2\,{ m_\sg}\,{ m_t}\,{ p_3q_2}\, \left(
    {{R^{(b)}_{1a}}^2}\,R^{(t)}_{1b}\, R^{(t)}_{2b}+
    {{R^{(b)}_{2a}}^2}\,R^{(t)}_{1b}\,R^{(t)}_{2b} \right) \\&&+ {{{
    m_\sg}}^2}\,{ p_3k}\, \left( {{R^{(b)}_{2a}}^2}\,{{R^{(t)}_{1b}}^2} +
    {{R^{(b)}_{1a}}^2}\,{{R^{(t)}_{2b}}^2} \right) \\&&- 2\,{ kq_2}\,{ m_b}\,{
    m_\sg}\, \left( R^{(b)}_{1a}\,R^{(b)}_{2a}\, {{R^{(t)}_{1b}}^2} +
    R^{(b)}_{1a}\,R^{(b)}_{2a}\,{{R^{(t)}_{2b}}^2} \right) \\&&\left.+ \left(
    2\,{ kq_2}\,{ p_3q_2} - { p_3k}\,{{{ q_2}}^2} \right) \, \left(
    {{R^{(b)}_{1a}}^2}\,{{R^{(t)}_{1b}}^2} +
    {{R^{(b)}_{2a}}^2}\,{{R^{(t)}_{2b}}^2} \right) \right)
\end{array}
$$
\vspace{0.75cm}
$$
\begin{array}{lcl}
 Q_{d_\alpha d_\beta}&=&{{3\,{g^4} }\over {4\,\left( {{{ q_2}}^2} -
   {{m_{{\chi^0}_\alpha}}^2} \right) \, \left( {{{ q_2}}^2} - {{m_{{\chi^0}_\beta}}^2}
   \right) }} \,\left( \phantom{+ { p_3k}\,m_{{\chi^0}_\beta}\,\left(
   {A_{-}}_{b\beta}^{(0)t} \right) } \right.\\&& {{{ q_2}}^2}\,{ m_b}\,{ m_t}\,
   \left( {A_{+}}_{a\beta}^{(0)b}\,{A_{+}}_{b\beta}^{(0)t}\,
   {A_{-}}_{a\alpha}^{(0)b}\,{A_{-}}_{b\alpha}^{(0)t} +
   {A_{-}}_{a\beta}^{(0)b}\,{A_{-}}_{b\beta}^{(0)t}\,
   {A_{+}}_{a\alpha}^{(0)b}\,{A_{+}}_{b\alpha}^{(0)t} \right) \\&&+ \left( 2\,{
   kq_2}\,{ p_3q_2} - { p_3k}\,{{{ q_2}}^2} \right) \, \left(
   {A_{+}}_{a\beta}^{(0)b}\,{A_{+}}_{b\beta}^{(0)t} \,
   {A_{+}}_{a\alpha}^{(0)b}\,{A_{+}}_{b\alpha}^{(0)t} +
   {A_{-}}_{a\beta}^{(0)b}\,{A_{-}}_{b\beta}^{(0)t}\,
   {A_{-}}_{a\alpha}^{(0)b}\,{A_{-}}_{b\alpha}^{(0)t} \right) \\&&+ { p_3q_2}\,{
   m_t}\, m_{{\chi^0}_\alpha} \, \left( {A_{+}}_{a\beta}^{(0)b}\,
   {A_{+}}_{b\beta}^{(0)t}\,{A_{+}}_{a\alpha}^{(0)b}\,{A_{-}}_{b\alpha}^{(0)t} +
   {A_{-}}_{a\beta}^{(0)b}\,{A_{-}}_{b\beta}^{(0)t}\,
   {A_{-}}_{a\alpha}^{(0)b}\,{A_{+}}_{b\alpha}^{(0)t} \right) \\&&+ { kq_2}\,{
   m_b}\,m_{{\chi^0}_\alpha}\,\left( {A_{+}}_{a\beta}^{(0)b}\,
   {A_{+}}_{b\beta}^{(0)t}\, {A_{-}}_{a\alpha}^{(0)b}\, {A_{+}}_{b\alpha}^{(0)t}+
   {A_{-}}_{a\beta}^{(0)b}\,{A_{-}}_{b\beta}^{(0)t}\,
   {A_{+}}_{a\alpha}^{(0)b}\,{A_{-}}_{b\alpha}^{(0)t} \right) \\&&+ { p_3q_2}\,{
   m_t}\,m_{{\chi^0}_\beta}\,\left({A_{-}}_{a\beta}^{(0)b}\, {A_{+}}_{b\beta}^{(0)t}\,
   {A_{-}}_{a\alpha}^{(0)b}\, {A_{-}}_{b\alpha}^{(0)t} +
   {A_{+}}_{a\beta}^{(0)b}\,{A_{-}}_{b\beta}^{(0)t}\,
   {A_{+}}_{a\alpha}^{(0)b}\,{A_{+}}_{b\alpha}^{(0)t} \right) \\&&+ { kq_2}\,{
   m_b}\,m_{{\chi^0}_\beta} \,\left(
   {A_{-}}_{a\beta}^{(0)b}\,{A_{+}}_{b\beta}^{(0)t}\,{A_{+}}_{a\alpha}^{(0)b}\,
   {A_{+}}_{b\alpha}^{(0)t} +
   {A_{+}}_{a\beta}^{(0)b}\,{A_{-}}_{b\beta}^{(0)t}\,{A_{-}}_{a\alpha}^{(0)b}\,
   {A_{-}}_{b\alpha}^{(0)t} \right) \\&&+ { m_b}\,{
   m_t}\,m_{{\chi^0}_\alpha}\,m_{{\chi^0}_\beta}\,\left(
   {A_{-}}_{a\beta}^{(0)b}\,{A_{+}}_{b\beta}^{(0)t}\,
   {A_{+}}_{a\alpha}^{(0)b}\,{A_{-}}_{b\alpha}^{(0)t} +
   {A_{+}}_{a\beta}^{(0)b}\,{A_{-}}_{b\beta}^{(0)t}\,
   {A_{-}}_{a\alpha}^{(0)b}\,{A_{+}}_{b\alpha}^{(0)t} \right) \\&&\left.+ {
   p_3k}\,m_{{\chi^0}_\alpha}\,m_{{\chi^0}_\beta}\,\left( {A_{-}}_{a\beta}^{(0)b}\,
   {A_{+}}_{b\beta}^{(0)t}\,{A_{-}}_{a\alpha}^{(0)b}\,{A_{+}}_{b\alpha}^{(0)t} +
   {A_{+}}_{a\beta}^{(0)b}\,{A_{-}}_{b\beta}^{(0)t}\, {A_{+}}_{a\alpha}^{(0)b}\,
   {A_{-}}_{b\alpha}^{(0)t} \right) \right)
\end{array}
$$
\vspace{0.75cm}
$$
\begin{array}{lcl}
  Q_{ab}+Q_{ab}^\dagger&=&{{3\,{g^4}\,\left( {{2\,\left( {{{ q_1}}^2} -{{{
         m_\hplus}}^2} \right) \, \left( {{{ q_1}}^2} -{{{ m_W}}^2} \right) }} +
         {{2\,{ \epsilon_\hplus}\,{ \epsilon_W}}} \right)
         \,R^{(b)}_{1a}\,R^{(t)}_{1b}\, G_{ab} }\over {4\,{{{ m_W}}^2}\, {\left(
         {{\left( {{{ q_1}}^2} -{{{ m_\hplus}}^2} \right) }^2} +{{{
         \epsilon_\hplus }}^2} \right) \,\left( {{\left( {{{ q_1}}^2} -{{{
         m_W}}^2} \right) }^2} + {{{ \epsilon_W}}^2} \right) } }}\\&& \,\left(
         {{{ m_t}}^2}\,{ \cot\beta}\, \left( { p_2p_3} -{ p_1p_3} + {{ {{{
         m_\st}}^2} -{{{ m_\sbt}}^2} }\over {{{{ m_W}}^2}}} \, { p_3q_1} \right)
         \right.\\&&\left. + { {{{ m_b}}^2}\,\tan\beta} \,\left( { p_2k}- {
         p_1k} + {{ {{{ m_\st}}^2} -{{{ m_\sbt}}^2} }\over {{{{ m_W}}^2}}}\,{
         kq_1} \right) \right)
\end{array}
$$
\vspace{0.75cm}
$$
\begin{array}{lcl}
 Q_{ac}+Q_{ac}^\dagger&=&{{{g^2}\,{{{ g_s}}^2\,\left( {{{ q_1}}^2} -{{{ m_W}}^2}
   \right) \, { Sum}\,R^{(b)}_{1a}\,R^{(t)}_{1b}\, }}\over {6\,\left( {{{
   {{\left( {{{ q_1}}^2} -{{{ m_W}}^2} \right) }^2} + \epsilon_W}}^2}\right) \,
   \left( {{{ q_2}}^2} -{{{ m_\sg}}^2} \right) }} \\&& \left( \left( {
   kq_2}\,\left( { p_2p_3} -{ p_1p_3} + {{{ {{{ m_\st}}^2} - {{{ m_\sbt}}^2}
   }\over {{{{ m_W}}^2}}} \, { p_3q_1}} \right) + \left( { p_2k} - { p_1k} +{{
   {{{ m_\st}}^2} - {{{ m_\sbt}}^2} } \over {{{{ m_W}}^2}}}\,{ kq_1} \right) \,
   { p_3q_2} \right.\right.\\&&\left.- { p_3k}\, \left( { p_2q_2} -{ p_1q_2} +
   {{ {{{ m_\st}}^2} - {{{ m_\sbt}}^2} }\over {{{{ m_W}}^2}}} \, { q_1q_2}
   \right) \right) \, R^{(b)}_{1a}\,R^{(t)}_{1b} \\&&- { m_b}\,{ m_\sg}\,\left(
   { p_2k} - { p_1k} + {{ {{{ m_\st}}^2} - {{{ m_\sbt}}^2} }\over {{{{
   m_W}}^2}}}\,{ kq_1}\right) \, R^{(b)}_{2a}\,R^{(t)}_{1b} \\&&- { m_\sg}\,{
   m_t}\,\left( { p_2p_3}-{ p_1p_3} + {{ {{{ m_\st}}^2} - {{{ m_\sbt}}^2}}\over
   {{{{ m_W}}^2}}} \,{ p_3q_1} \right) \,R^{(b)}_{1a}\,R^{(t)}_{2b} \\&&\left.+
   { m_b}\,{ m_t}\,\left( { p_2q_2} -{ p_1q_2} + {{ {{{ m_\st}}^2} - {{{
   m_\sbt}}^2} }\over {{{{ m_W}}^2}}} \,{ q_1q_2} \right)
   \,R^{(b)}_{2a}\,R^{(t)}_{2b} \right)
\end{array}
$$
\vspace{0.75cm}
$$
\begin{array}{lcl}
Q_{ad}+Q_{ad}^\dagger&=&{{3\,{g^4}\,\left({{{ q_1}}^2} -{{{ m_W}}^2}\right)\,
   R^{(b)}_{1a}\,R^{(t)}_{1b} }\over {2\,\left( {{\left( {{{ q_1}}^2} -{{{
   m_W}}^2} \right) }^2} +{{{ \epsilon_W}}^2} \right) \, \left( {{{ q_2}}^2} -
   {{m_{{\chi^0}_\alpha}}^2} \right) }}\, \left( \phantom{ { m_t}\,\left( { p_2p_3}
   {{\left( {{{ m_\sbt}}^2} \right) \,{ p_3q_1}}\over {{{{ m_W}}^2}}} \right)
   {A_{-}}_{b\alpha}^{(0)t}\,m_{{\chi^0}_\alpha}} \right.\\&& \left( { kq_2}\,\left(
   { p_2p_3} -{ p_1p_3} + {{ {{{ m_\st}}^2} - {{{ m_\sbt}}^2}}\over {{{{
   m_W}}^2}}} \, { p_3q_1} \right) + \left( {{ {{{ m_\st}}^2} - {{{ m_\sbt}}^2}}
   \over {{{{ m_W}}^2}}}\,{ kq_1} - { p_1k} + { p_2k} \right) \, { p_3q_2}
   \right.\\&&\left. - { p_3k}\, \left( { p_2q_2} -{ p_1q_2} + {{ {{{ m_\st}}^2}
   - {{{ m_\sbt}}^2} }\over {{{{ m_W}}^2}}} \, { q_1q_2} \right) \right) \,
   {A_{+}}_{a\alpha}^{(0)b}\,{A_{+}}_{b\alpha}^{(0)t} \\&&+ { m_b}\,{ m_t}\,\left( {
   p_2q_2} -{ p_1q_2} +{{ {{{ m_\st}}^2} - {{{ m_\sbt}}^2} }\over {{{{
   m_W}}^2}}}\,{ q_1q_2} \right) \,{A_{-}}_{a\alpha}^{(0)b}\,
   {A_{-}}_{b\alpha}^{(0)t} \\&& + { m_b}\, m_{{\chi^0}_\alpha}\,\left({ p_2k} - {
   p_1k} + {{ {{{ m_\st}}^2} - {{{ m_\sbt}}^2} }\over {{{{ m_W}}^2}}}\, {
   kq_1}\right)\, {A_{-}}_{a\alpha}^{(0)b} \, {A_{+}}_{b\alpha}^{(0)t} \\&& \left.+
   { m_t}\,m_{{\chi^0}_\alpha}\,\left( -{ p_1p_3} + { p_2p_3} + {{ {{{ m_\st}}^2} -
   {{{ m_\sbt}}^2} }\over {{{{ m_W}}^2}}} \,{ p_3q_1} \right)
   \,{A_{+}}_{a\alpha}^{(0)b}\, {A_{-}}_{b\alpha}^{(0)t} \right)
\end{array}
$$
\vspace{0.75cm}
$$
\begin{array}{lcl}
Q_{bc}+Q_{bc}^\dagger&=& {{{g^2}\,{{{ g_s}}^2}\,\left( {{{ q_1}}^2}-{{{
     m_\hplus}}^2} \right) \, { Sum}\, G_{ab} }\over {6\,{{{ m_W}}^2}\,\left(
     {{\left( {{{ q_1}}^2} -{{{ m_\hplus}}^2} \right) }^2}+ {{{
     \epsilon_\hplus}}^2} \right) \left( {{{ q_2}}^2}-{{{ m_\sg}}^2} \right)
     }}\, \left(\phantom{{ p_3q_2}\,\left( { m_b}\,{ \tan\beta}\,R^{(b)}_{2a}
     \right)} \right.\\&&{ m_\sg}\,{ p_3k}\, \left( - { m_b}\,{
     \tan\beta}\,R^{(b)}_{2a}\, R^{(t)}_{1b} - { m_t}\,{
     \cot\beta}\,R^{(b)}_{1a}\,R^{(t)}_{2b} \right) \\&&+ { m_b}\,{ m_\sg}\,{
     m_t}\, \left( - { m_t}\,{ \cot\beta}\,R^{(b)}_{2a}\, R^{(t)}_{1b} - {
     m_b}\,{ \tan\beta}\,R^{(b)}_{1a}\,R^{(t)}_{2b} \right) \\&&+ { m_b}\,{
     kq_2}\,\left( { m_b}\,{ \tan\beta}\,R^{(b)}_{1a}\, R^{(t)}_{1b} + { m_t}\,{
     \cot\beta}\,R^{(b)}_{2a}\,R^{(t)}_{2b} \right) \\&&\left.+ { m_t}\,{
     p_3q_2}\,\left( { m_t}\,{ \cot\beta}\,R^{(b)}_{1a}\, R^{(t)}_{1b} + {
     m_b}\,{ \tan\beta}\,R^{(b)}_{2a}\,R^{(t)}_{2b} \right) \right)
\end{array}
$$
\vspace{0.75cm}
$$
\begin{array}{lcl}
Q_{bd}+Q_{bd}^\dagger&=&{{3\,{g^4}\,\left( {{{ q_1}}^2} -{{{ m_\hplus}}^2}
   \right)\, G_{ab} }\over {2\,{{{ m_W}}^2}\,\left( {{{ {{\left( {{{ q_1}}^2}
   -{{{ m_\hplus}}^2} \right) }^2} + \epsilon_\hplus}}^2}\right) \, \left( {{{
   q_2}}^2} - {{m_{{\chi^0}_\alpha}}^2} \right) }}\left(
   \phantom{m_{{\chi^0}_\alpha}\,\left( { m_b}\,{ \tan\beta} {A_{-}}_{b\alpha}^{(0)t}
   \right)} \right.\\&&{ m_b}\,{ kq_2}\, \left( { m_b}\,{
   \tan\beta}\,{A_{+}}_{a\alpha}^{(0)b}\, {A_{+}}_{b\alpha}^{(0)t} + { m_t}\,{
   \cot\beta}\,{A_{-}}_{a\alpha}^{(0)b}\, {A_{-}}_{b\alpha}^{(0)t} \right) \\&&+ {
   m_t}\,{ p_3q_2}\,\left( { m_t}\,{ \cot\beta}\,
   {A_{+}}_{a\alpha}^{(0)b}\,{A_{+}}_{b\alpha}^{(0)t} + { m_b}\,{
   \tan\beta}\,{A_{-}}_{a\alpha}^{(0)b}\, {A_{-}}_{b\alpha}^{(0)t} \right) \\&&+
   m_{{\chi^0}_\alpha}\, { p_3k}\,\left( { m_b}\,{
   \tan\beta}\,{A_{-}}_{a\alpha}^{(0)b}\, {A_{+}}_{b\alpha}^{(0)t} + { m_t}\,{
   \cot\beta}\,{A_{+}}_{a\alpha}^{(0)b}\, {A_{-}}_{b\alpha}^{(0)t} \right)
   \\&&\left.+ { m_b}\,{ m_t}\,m_{{\chi^0}_\alpha}\,\left( { m_t}\,{
   \cot\beta}\,{A_{-}}_{a\alpha}^{(0)b}\, {A_{+}}_{b\alpha}^{(0)t} + { m_b}\,{
   \tan\beta}\,{A_{+}}_{a\alpha}^{(0)b}\, {A_{-}}_{b\alpha}^{(0)t} \right) \right)
\end{array}
$$
\vspace{0.75cm}
$$
\begin{array}{lcl}
Q_{cd}&=&{{{g^2}\,{{{ g_s}}^2}\,{ Sum}}\over {12\,\left( {{{ q_2}}^2} -{{{
   m_\sg}}^2} \right) \, \left( {{{ q_2}}^2} - {{m_{{\chi^0}_\beta}}^2} \right)
   }}\, \left( \phantom{+ \left( { p_3k}\,{{{ q_2}}^2} \right) \, \left(
   {A_{-}}_{a\beta}^{(0)b} R^{(t)}_{2b} \right) } \right.\\&& -{ m_\sg}\,{ m_t}\,{
   p_3q_2}\, \left( {A_{-}}_{a\beta}^{(0)b}\,
   {A_{-}}_{b\beta}^{(0)t}\,R^{(b)}_{2a}\, R^{(t)}_{1b} +
   {A_{+}}_{a\beta}^{(0)b}\,{A_{+}}_{b\beta}^{(0)t}\, R^{(b)}_{1a}\,R^{(t)}_{2b}
   \right) \\&&- { m_b}\,{ m_\sg}\,{ m_t}\,m_{{\chi^0}_\beta}\, \left(
   {A_{+}}_{a\beta}^{(0)b}\, {A_{-}}_{b\beta}^{(0)t}\,R^{(b)}_{2a}\, R^{(t)}_{1b} +
   {A_{-}}_{a\beta}^{(0)b}\,{A_{+}}_{b\beta}^{(0)t}\, R^{(b)}_{1a}\,R^{(t)}_{2b}
   \right) \\&&- { m_\sg}\,m_{{\chi^0}_\beta}\,{ p_3k}\, \left(
   {A_{-}}_{a\beta}^{(0)b}\,{A_{+}}_{b\beta}^{(0)t}\,R^{(b)}_{2a}\,R^{(t)}_{1b} +
   {A_{+}}_{a\beta}^{(0)b}\, {A_{-}}_{b\beta}^{(0)t}\,R^{(b)}_{1a}\,R^{(t)}_{2b}
   \right) \\&&- { m_b}\,{ m_\sg}\,{ kq_2}\, \left( {A_{+}}_{a\beta}^{(0)b}\,
   {A_{+}}_{b\beta}^{(0)t}\,R^{(b)}_{2a}\, R^{(t)}_{1b} +
   {A_{-}}_{a\beta}^{(0)b}\,{A_{-}}_{b\beta}^{(0)t}\, R^{(b)}_{1a}\,R^{(t)}_{2b}
   \right) \\&&+ { m_b}\,{ m_t}\,{{{ q_2}}^2}\, \left(
   {A_{-}}_{a\beta}^{(0)b}\,{A_{-}}_{b\beta}^{(0)t}\, R^{(b)}_{1a}\,R^{(t)}_{1b} +
   {A_{+}}_{a\beta}^{(0)b}\,{A_{+}}_{b\beta}^{(0)t}\, R^{(b)}_{2a}\,R^{(t)}_{2b}
   \right) \\&&+ { m_t}\,,m_{{\chi^0}_\beta}\,{ p_3q_2}\ \left(
   {A_{+}}_{a\beta}^{(0)b}\,{A_{-}}_{b\beta}^{(0)t}\, R^{(b)}_{1a}\,R^{(t)}_{1b} +
   {A_{-}}_{a\beta}^{(0)b}\,{A_{+}}_{b\beta}^{(0)t}\, R^{(b)}_{2a}\,R^{(t)}_{2b}
   \right) \\&&+ { m_b}\,m_{{\chi^0}_\beta}\,{ kq_2}\, \left(
   {A_{-}}_{a\beta}^{(0)b}\,{A_{+}}_{b\beta}^{(0)t}\, R^{(b)}_{1a}\,R^{(t)}_{1b} +
   {A_{+}}_{a\beta}^{(0)b}\,{A_{-}}_{b\beta}^{(0)t}\, R^{(b)}_{2a}\,R^{(t)}_{2b}
   \right) \\&&\left.+ \left( 2\,{ kq_2}\,{ p_3q_2} - { p_3k}\,{{{ q_2}}^2}
   \right) \, \left( {A_{+}}_{a\beta}^{(0)b}\,{A_{+}}_{b\beta}^{(0)t}\,
   R^{(b)}_{1a}\,R^{(t)}_{1b} + {A_{-}}_{a\beta}^{(0)b}\,{A_{-}}_{b\beta}^{(0)t}\,
   R^{(b)}_{2a}\,R^{(t)}_{2b} \right) \right)
\end{array}
$$

\vspace{0.7cm}

The decay rate is obtained from the formulae above by converting all scalar products
$p_ip_j\equiv p_i.p_j$  involved in these equations in terms of Mandelstam
invariants $S_1,S_2,S_3$ and then by performing integration over them in the
standard manner\,\cite{BK}. 
Care has to be exercised in the numerical
integrations near the poles.

\vspace{1cm}

\newpage
\vspace{1cm}
\begin{center}
\begin{Large}
{\bf Figure Captions}
\end{Large}
\end{center}
\begin{itemize}
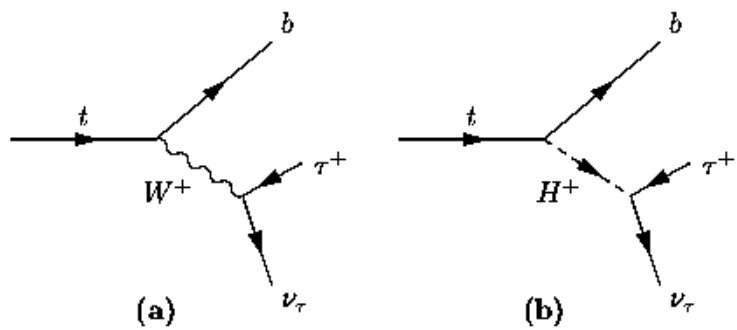
\item{\bf Fig.1} Feynman diagrams for the decay {\bf I}. Diagram {\bf (a)} is the SM 
amplitude, and diagram {\bf (b)} is the extra charged Higgs contribution in the
MSSM.

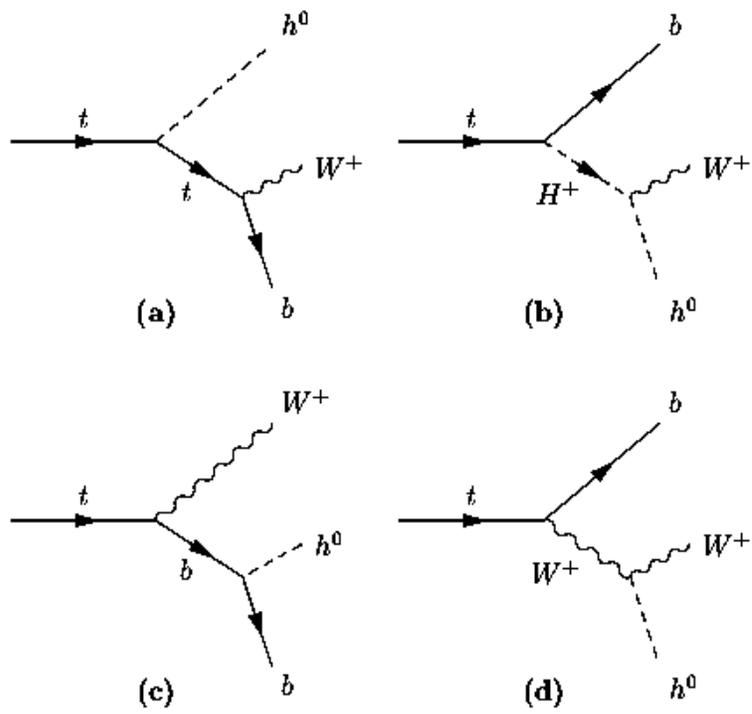
\item{\bf Fig.2} Feynman diagrams for the decay {\bf II}.

\item{\bf Fig.3} Feynman diagrams for the decay {\bf III}.

\item{\bf Fig.4} Feynman diagrams for the decay {\bf IV}.

\item{\bf Fig.5} Feynman diagrams for the decay {\bf V}.

\item{\bf Fig.6} Feynman diagrams for the decays {\bf VI} {\bf (a)} 
and {\bf VII} {\bf (b)}.

\item{\bf Fig.7} Feynman diagrams for the decay {\bf VIII}.

\item{\bf Fig.8} Feynman diagrams for the decay {\bf IX}. Those
for the decay {\bf X} are obtained from
{\bf (a)} and {\bf (b)} after replacing $\st \rightarrow \sneut_\tau$ 
and $\sbt \rightarrow \stau$.

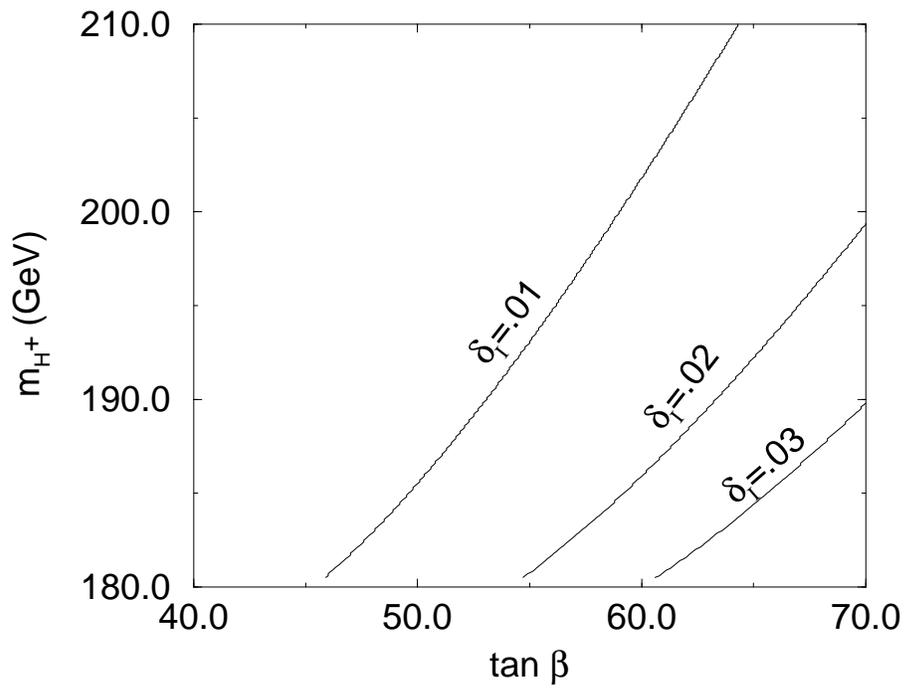
\item{\bf Fig. 9} Isolines of $\delta_I$ in the $\tan\beta$-$m_\hplus$ plane.

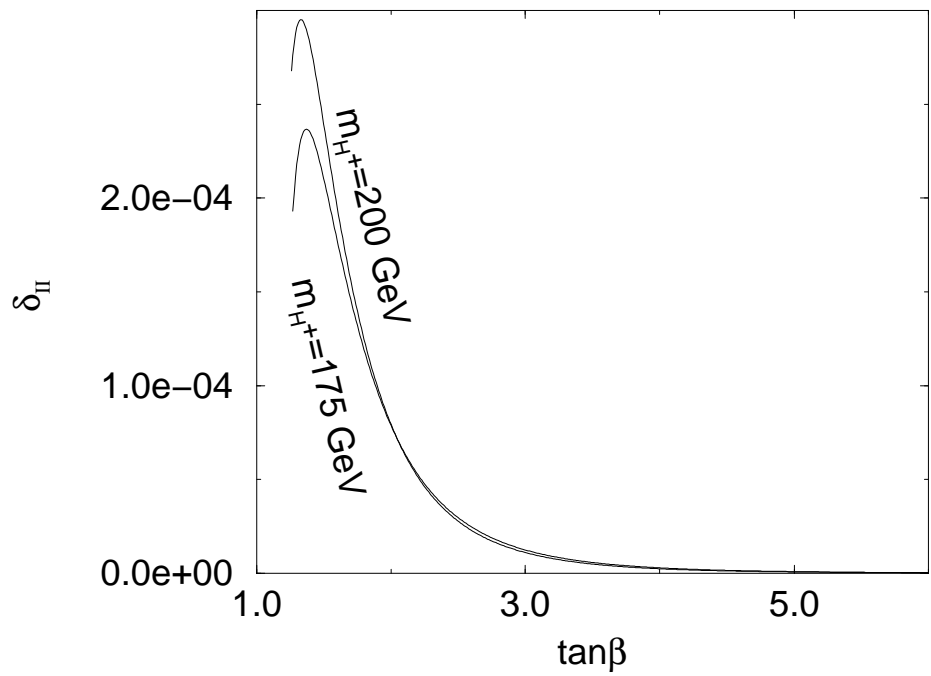
\item{\bf Fig. 10} $\delta_{II}$ as a function of $\tan\beta$ for two values of
$m_\hplus$.

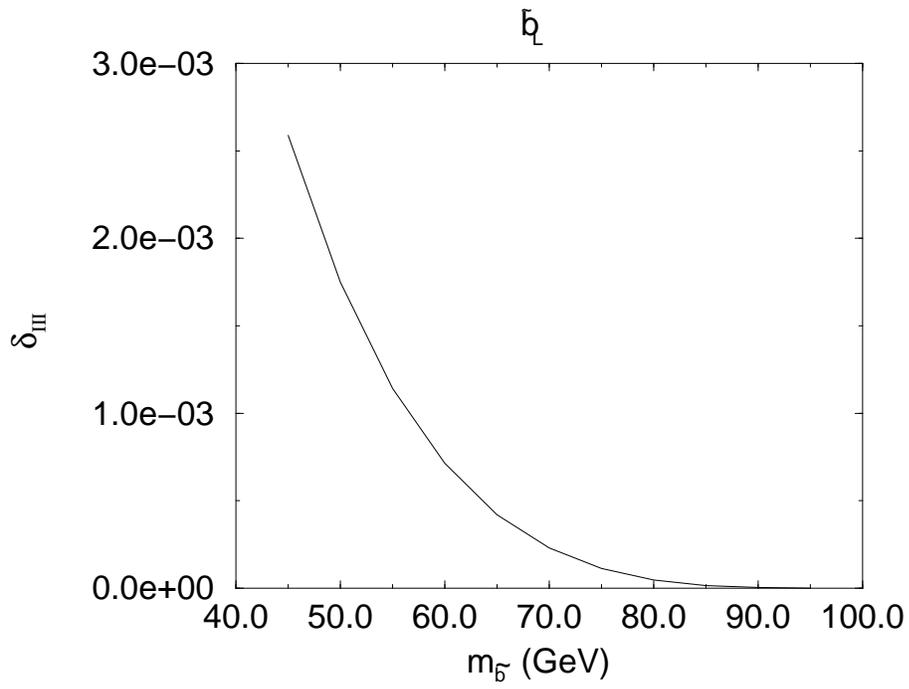
\item{\bf Fig. 11} $\delta_{III}$ as a function of $m_\sbt$ for
a $\sbt_L$ final state,
with $\tan\beta=1$ and $m_\sg=1 \GeV$. 

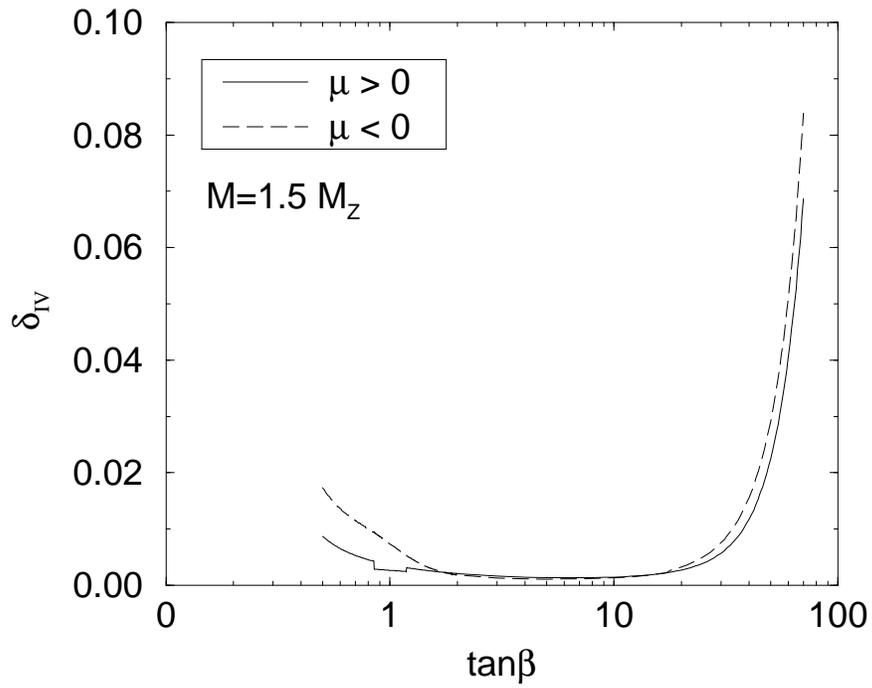
\item{\bf Fig. 12}  Maximum value of $\delta_{IV}$ as a function of
$\tan\beta$, for $m_\sbt=125 \GeV$, $m_\st=155\GeV$,
$m_\hplus=175\GeV$ and $m_\cplus = 60 \GeV$. 

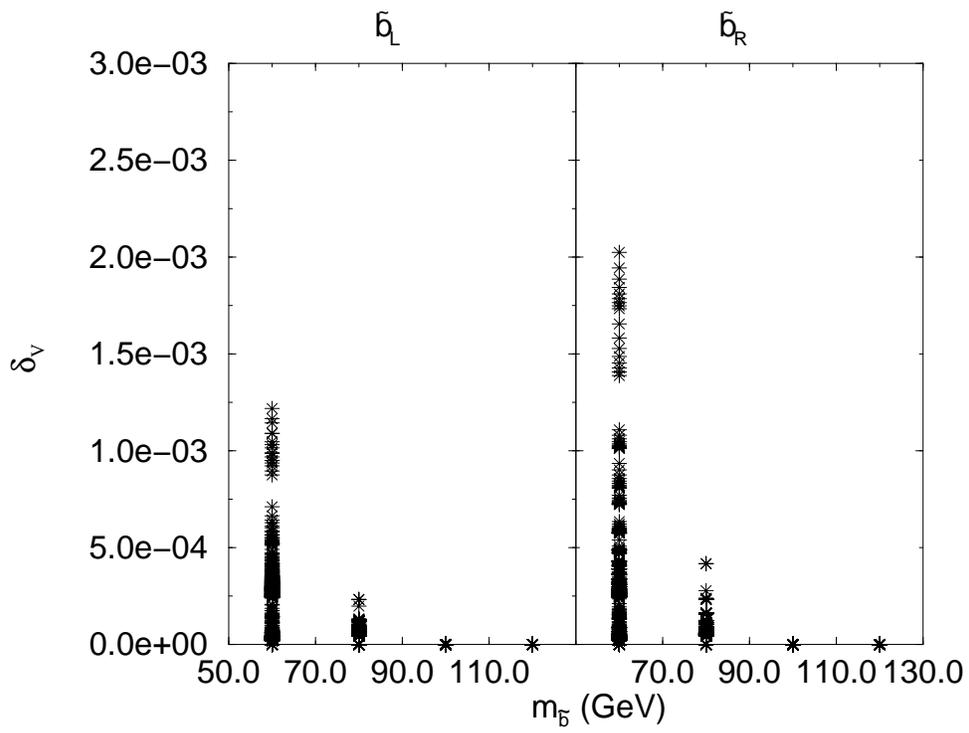
\item{\bf Fig. 13} Lattice plot sampling of $\delta_V$ for $\sbt_L$
and $\sbt_R$ final states as a function of $m_\sbt$, with $\tan\beta=40$,
$m_\sg=140 \GeV$, $-5 M_Z < \mu < 5 M_Z$,
$0 < M <5 M_Z$, and $125\GeV < M^t_{LR} < 230 \GeV$. In this case we have used 
$m_\cplus >50 \GeV$.

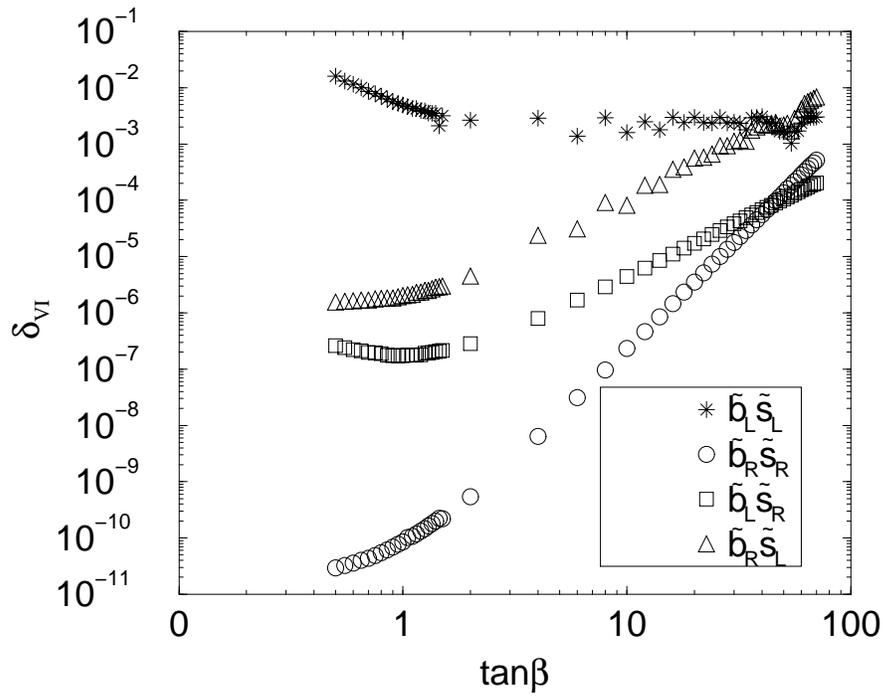
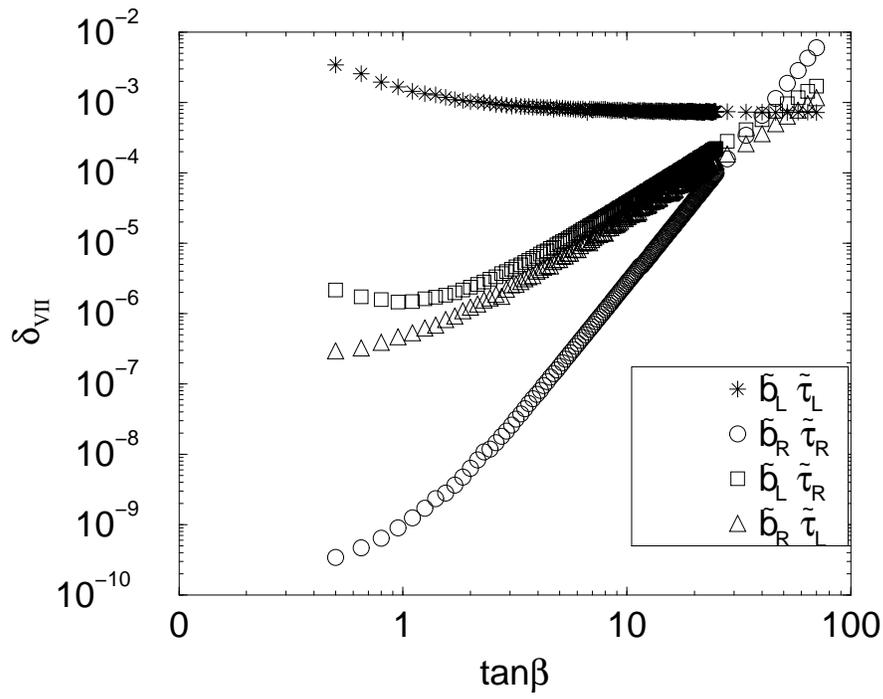
\item{\bf Fig. 14} {\bf (a)} Maximum value of $\delta_{VII}$ as a function of
$\tan\beta$ for $m_\cplus = 135 \GeV$ and $m_\sbt=m_{\tilde s}=50 \GeV$; 
{\bf (b)} As before, but for $\delta_{VIII}$ and 
 $m_\stau=95 \GeV$.

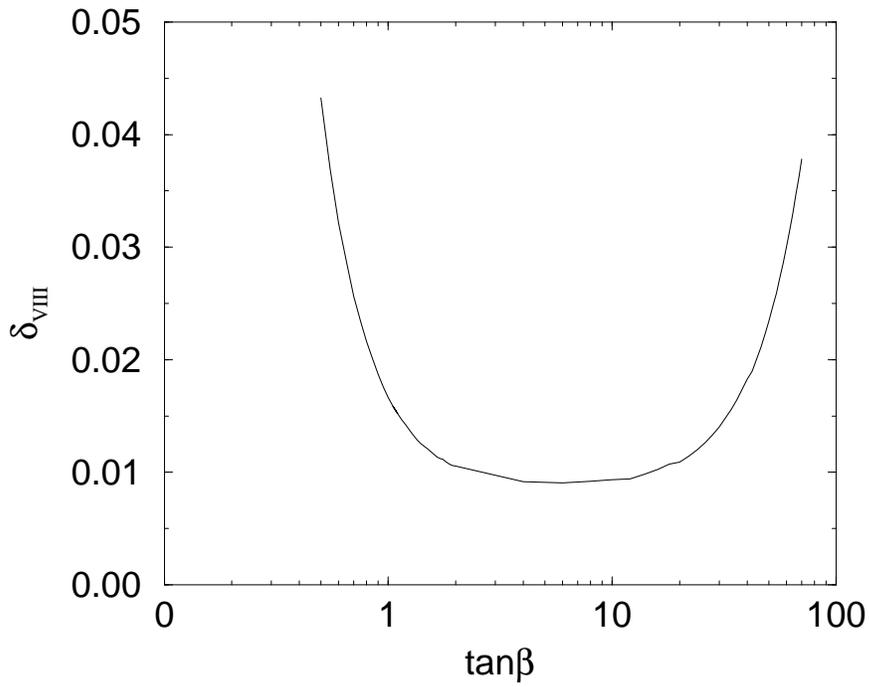
\item{\bf Fig. 15} Maximum value of $\delta_{VIII}$ for fixed 
$m_\sg=1\GeV$, $m_\st=190\GeV$
and $m_\sbt=150\GeV$. The maximum is attained at points on the $(\mu,M)$-plane 
corresponding to the phenomenological limit $m_\cplus = 60 \GeV$.

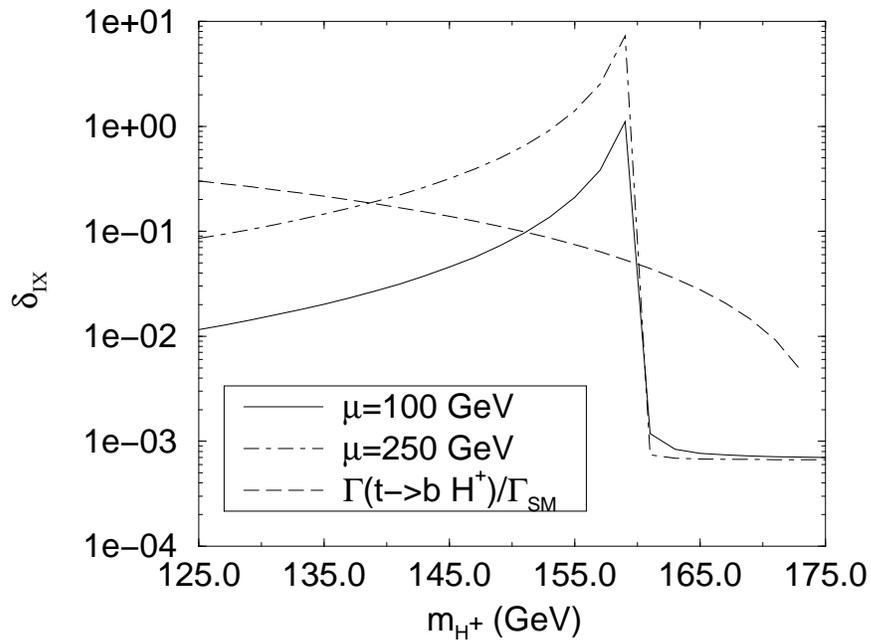
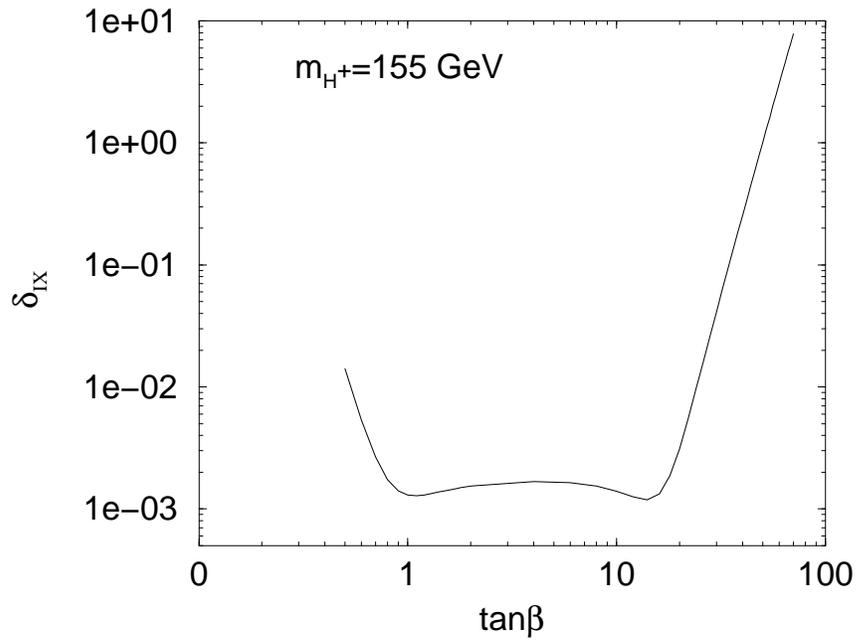
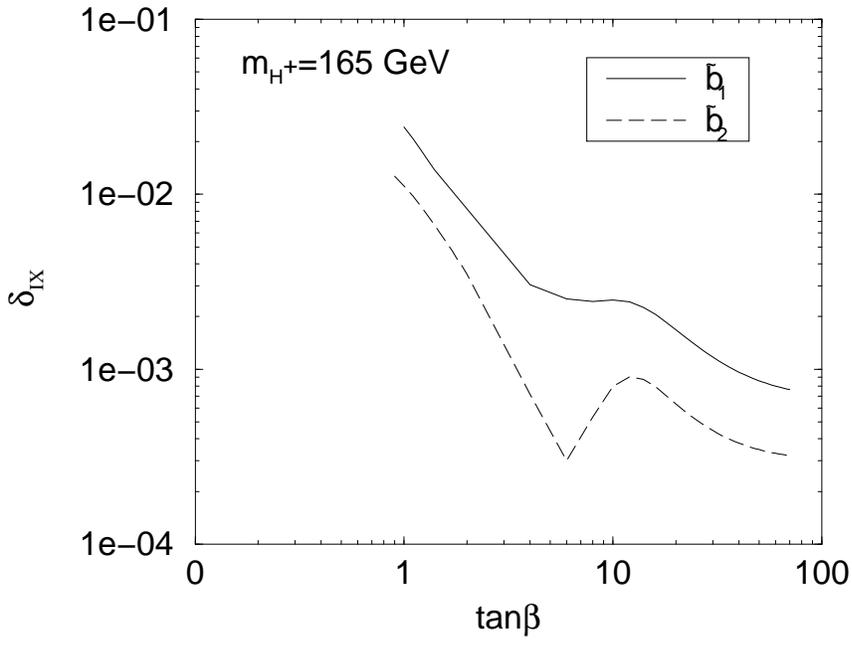
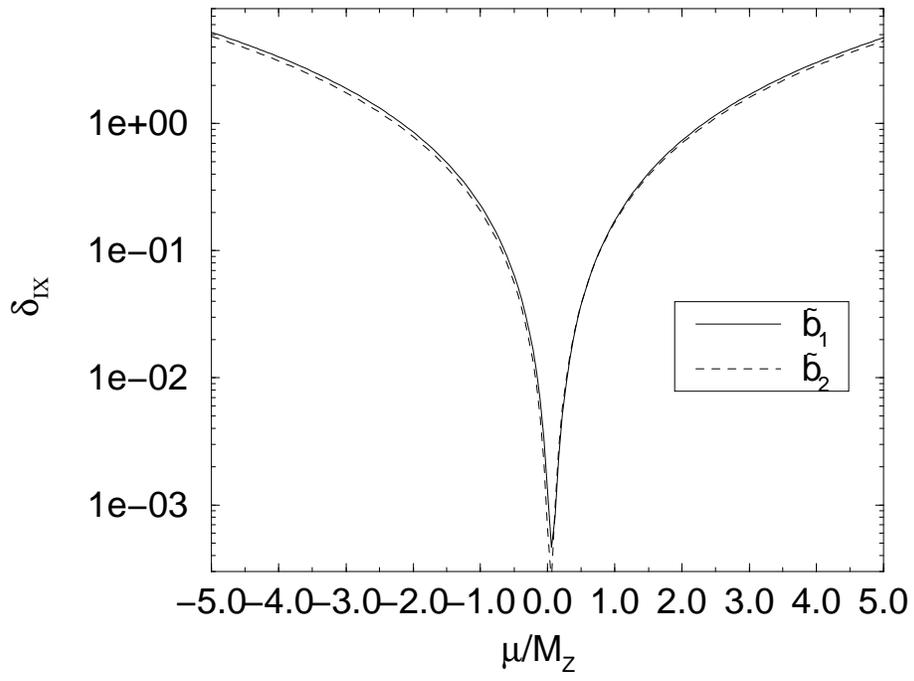
\item{\bf Fig. 16} {\bf (a)} $\delta_{IX}$ as a function of $m_\hplus$ for $\tan\beta=36$,
$m_\st=60\GeV$, $m_\sbt=100\GeV$, $m_\sg=130 \GeV$ and $\mu=100\GeV,250\GeV$. Also 
plotted is the ratio $\Gamma(t \rightarrow b \hplus)/\Gamma_{SM}$; 
{\bf (b)} $\delta_{IX}$ versus $\tan \beta$ for  $\mu=100\GeV$ and a 
charged Higgs mass
value below the  $\tilde{t}\,\bar{\tilde{b}}$ threshold;
{\bf (c)} As in (b) but for a charged Higgs mass above the 
$\tilde{t}\,\bar{\tilde{b}}$ threshold; {\bf (d)} $\delta_{IX}$ as a
function of $\mu$, for the same inputs as in (b) and $\tan\beta=36$.

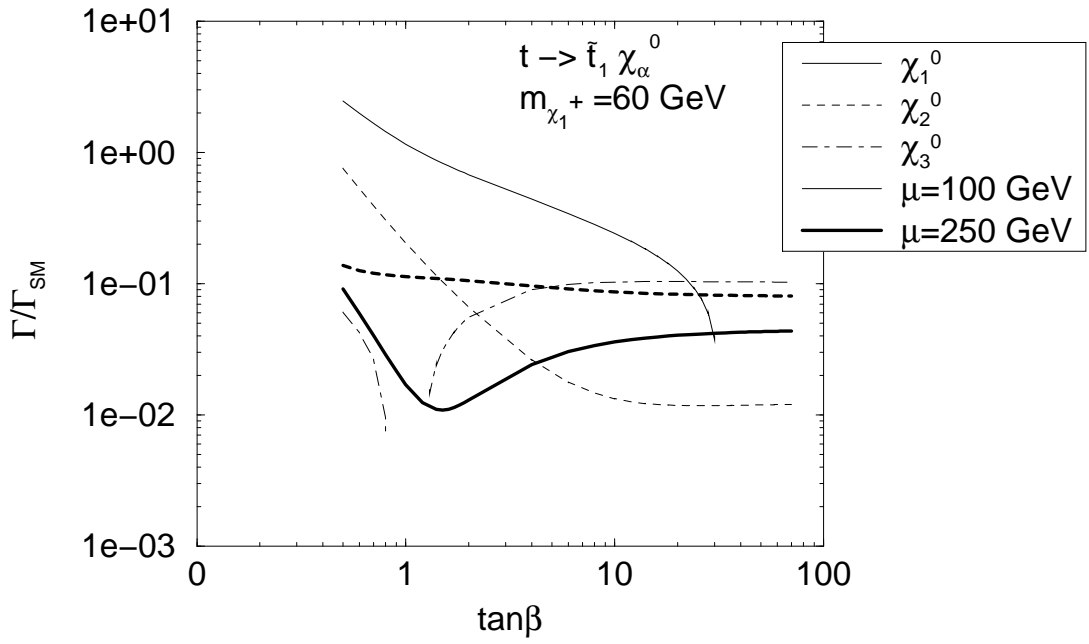
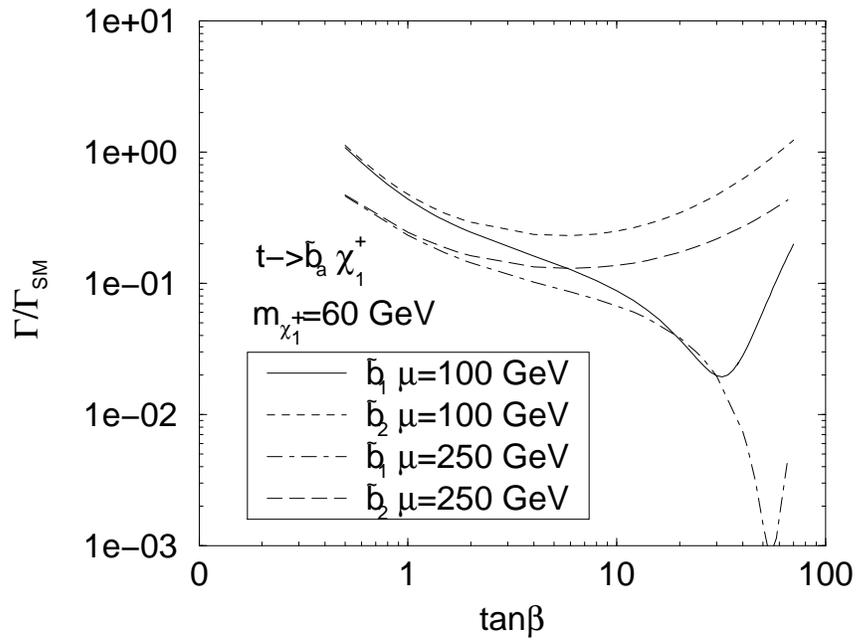
\item{\bf Fig. 17} {\bf (a)} $\Gamma(t \rightarrow \st \, \chi^0_\alpha)/ 
\Gamma_{SM}$ for the first three lightest neutralinos 
as a function of $\tan\beta$ for $m_\cplus=60 \GeV$, $m_\st=60\GeV$ 
and $\mu=100\GeV,250\GeV$; {\bf (b)} As before, but for the decay 
$t \rightarrow \sbt_a\, \cplus$ and $m_\sbt=100\GeV$.

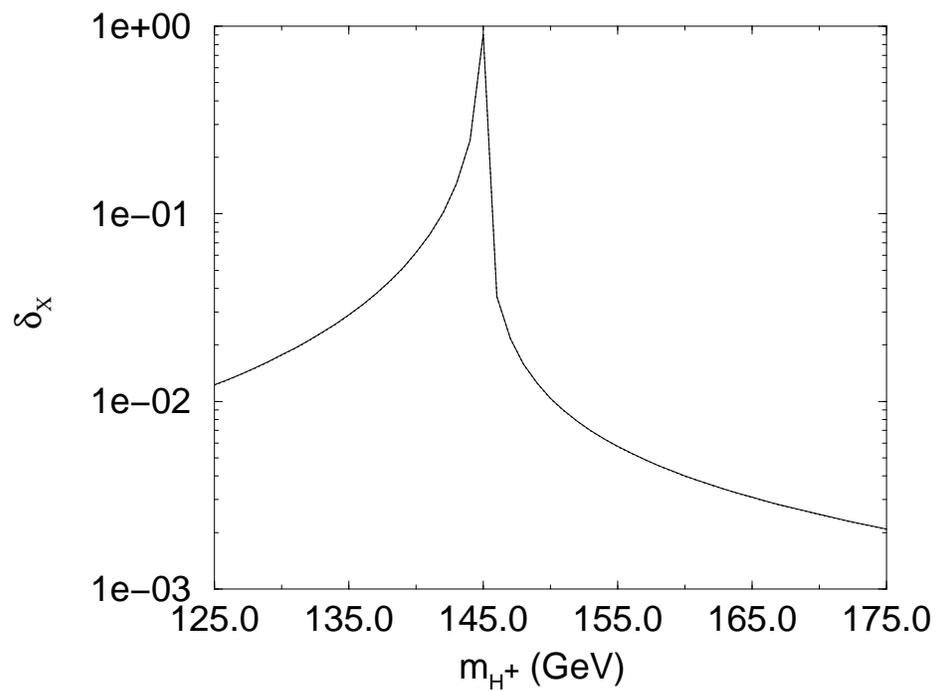
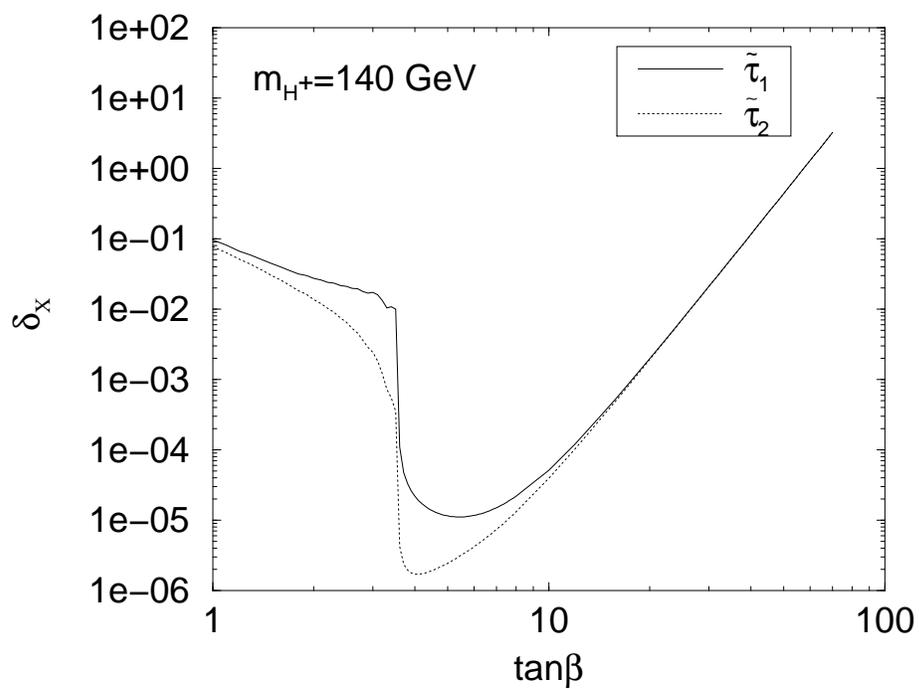
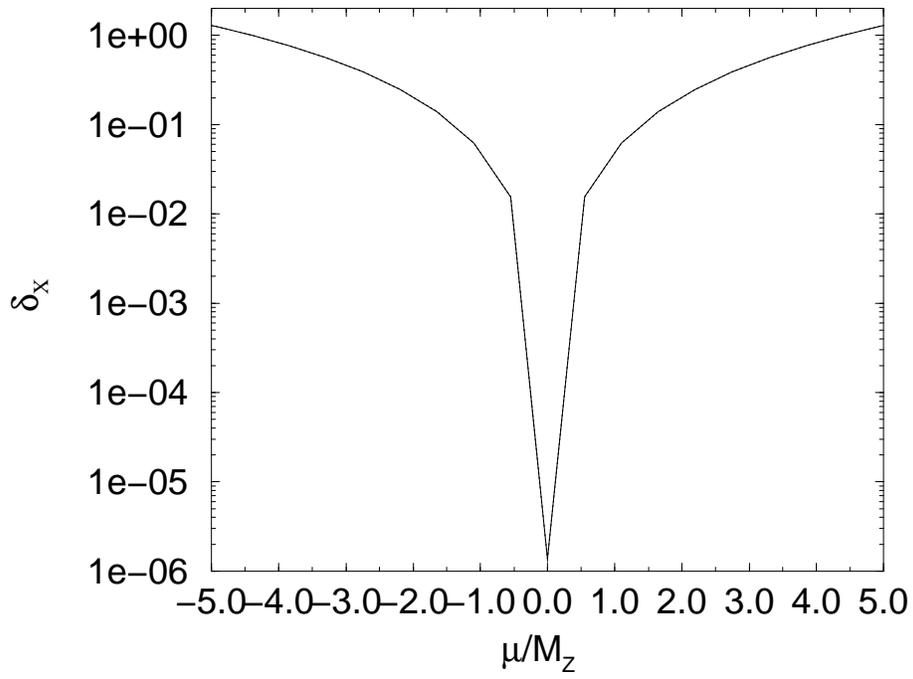
\item{\bf Fig. 18} {\bf (a)} $\delta_X$ as a function of $m_\hplus$ for $\tan\beta=36$,
$m_\sneut=50\GeV$ and $\mu=100\GeV$; {\bf (b)}  $\delta_X$
versus $\tan\beta$ for $m_\hplus=140\GeV$; {\bf (c)} $\delta_X$ as a function
of $\mu$ for the same inputs as before.

\end{itemize}

\vspace{2cm}

\newcommand{\neutdos}{\chi^0_2}

\begin{table}[htbp]
  \footnotesize
  \begin{center}
    \leavevmode
\begin{tabular}{ll}
\hline
Typical Decays & Signal \\
\hline
 (a) $\left\{ \begin{array}{l} \st_1 \rightarrow c \neut \\ \bar\sbt_{1,2}
  \rightarrow \bar b \neut \end{array} \right.$ &
  \parbox{7cm}{\begin{flushleft}2~j's~+~$\slas{p}$

(more $\slas{p}$, less $b$ activity)\end{flushleft}} \\

(b) $\left\{ \begin{array}{l} \st_1 \rightarrow c \neut \\ \bar\sbt_{1,2}
 \rightarrow \bar b \neutdos \rightarrow\left[\begin{array}{l}\bar b (Z^{(*)})
 \neut \rightarrow \bar b (f \bar f) \neut\\ {\rm or}\\ \bar b (h^0 \neut)
 \rightarrow \bar b (b \bar b \mbox{ or } \tau^+ \tau^-) \neut
 \end{array}\right.\end{array}\right.$ &\parbox{7cm}{\begin{flushleft}
 4~j's~+~$\slas{p}$; 2~j's~+~$\slas{p}$; 2~j's~+~$l^+\, l^-$~+~$\slas{p}$.

($Z^0$ or $h^0$ emission; 

more $b$ activity if $h^0$ or $Z^0\rightarrow b \bar b$; 

less $b$ activity if $h^0$ or $Z^0\rightarrow f \bar f$ or $\nu \bar \nu $)\end{flushleft}}\\
 (c) $\left\{ \begin{array}{l} \st_1 \rightarrow c \neut \\ \bar\sbt_{2}
  \rightarrow \bar c \cplus \rightarrow \bar c (f \bar f' \neut) \end{array}
  \right.$ & \parbox{7cm}{\begin{flushleft}4~j's~+~$\slas{p}$;
  2~j's~+~$l^+$~+~$\slas{p}$.

(isolated lepton $l^+$) \end{flushleft}}\\
(d) $\left\{ \begin{array}{l} \st_1 \rightarrow b \cplus \rightarrow b (f \bar
  f' \neut)\\ \bar\sbt_{1,2} \rightarrow \bar b \, \neut \end{array} \right.$ &
  \parbox{7cm}{\begin{flushleft}4~j's~+~$\slas{p}$; 2~j's~+~$l^+$~+~$\slas{p}$.

(more $\slas{p}$, more $b$ activity)\end{flushleft}} \\

 (e) $\left\{ \begin{array}{l} \st_1 \rightarrow b \cplus \rightarrow b (f \bar
  f' \neut)\\ \bar\sbt_{1,2} \rightarrow \bar b \, \neutdos \rightarrow \left[
  \begin{array}{l} \bar b(Z^{(*)}) \neut)\rightarrow \bar b (f \bar f) \neut \\
  \mbox{ or } \\ \bar b (h^0 \neut) \rightarrow \bar b (b \bar b \mbox{ or }
  \tau^+ \tau^-) \neut \end{array}\right.  \end{array} \right.$&
  \parbox{7cm}{\begin{flushleft}6~j's~+~$\slas{p}$;
  4~j's~+~$l^+\,l^-$~+~$\slas{p}$; 4~j's~+~$l^+$~+~$\slas{p}$;
  4~j's~+~$\slas{p}$; 2~j's~+~$l^+l^-\,l'^+$~+~$\slas{p}$;
  2~j's~+~$l^+$~+~$\slas{p}$.

($Z^0$ or $h^0$ emission;

more $b$ activity if $h^0$ or $Z^{*}\rightarrow b \bar b$; 

Trilepton)\end{flushleft}}\\

 (f) $\left\{ \begin{array}{l} \st_1 \rightarrow b \cplus \rightarrow b (f \bar
  f' \neut)\\ \bar\sbt_{1,2} \rightarrow \bar c \cplus \rightarrow \bar c (f
  \bar f' \neut)\end{array} \right.$&
  \parbox{7cm}{\begin{flushleft}6~j's~+~$\slas{p}$; 4~j's~+~$l^+$~+~$\slas{p}$;
  2~j's~+~$l^+\,l'^+$~+~$\slas{p}$.

(dilepton of same sign)\end{flushleft}}\\
\hline\end{tabular}
\caption{Detection signatures for the decay {\bf IX}. The decay chains
  (a)-(c) are possible only if $m_{\st_1} < m_{\cplus,\tilde l,\tilde \nu,\sbt}$
  and $m_{\neut} < m_{\st_1}$. The decay chains (d)-(f) are possible only if
  $m_{\st_1} > m_{\cplus}$. Here $\slas{p}$,j,$l^\pm$,$Z^{(*)}$ and $f$ denote
  missing energy-momentum, jet, isolated charged lepton, $Z^0$ real or virtual,
  and $(q,l^\pm,\nu)$, respectively.}
    \label{tab:canals}
  \end{center}
\normalsize
\end{table}

\begin{table}[htbp]
  \footnotesize
  \begin{center}
    \leavevmode
\begin{tabular}{ll}
\hline
Typical Decays & Signal \\
\hline
 (a) $\left\{ \begin{array}{l} \sneut_\tau \rightarrow \nu_\tau \neut \\
  \stau_{1,2}^+ \rightarrow \tau^+ \neut \end{array} \right.$
  &\parbox{7cm}{\begin{flushleft} $\tau^+$~+~$\slas{p}$

(more $\slas{p}$)\end{flushleft}} \\

(b) $\left\{ \begin{array}{l} \sneut_\tau \rightarrow \nu_\tau \neut \\
 \stau_{1,2}^+ \rightarrow \tau^+ \neutdos \rightarrow\left[\begin{array}{l}
 \tau^+ (Z^{(*)}) \neut \rightarrow \tau^+ (f \bar f) \neut\\ {\rm or}\\ \tau^+
 (h^0 \neut) \rightarrow \tau^+ (b \bar b \mbox{ or } \tau^+ \tau^-) \neut
 \end{array}\right.\end{array}\right.$ &
 \parbox{7cm}{\begin{flushleft}2~j's~+~$\tau^+$~+~$\slas{p}$; $\tau^+$~+~$l^+\,
 l^-$~+~$\slas{p}$.

($Z^0$ or $h^0$ emission; 

more $b$ activity if $h^0$ or $Z^0\rightarrow b \bar b$; 

less  $b$ activity if $h^0$ or $Z^0\rightarrow f \bar f$ or $\nu \bar \nu $)\end{flushleft}}\\
 (c) $\left\{ \begin{array}{l} \sneut_\tau \rightarrow \nu_\tau \neut \\
  \stau_{1,2}^+ \rightarrow \bar\nu_\tau \cplus \rightarrow \nu_\tau (f \bar f'
  \neut) \end{array} \right.$ &
  \parbox{7cm}{\begin{flushleft}2~j's~+~$\slas{p}$; $l^+$~+~$\slas{p}$.

(isolated lepton;

less $b$ activity) \end{flushleft}}\\
 (d) $\left\{ \begin{array}{l} \sneut_\tau \rightarrow \tau^- \cplus \rightarrow
  \tau^- (f \bar f' \neut)\\ \stau_{1,2}^+ \rightarrow \tau^+ \, \neut
  \end{array} \right.$ & \parbox{7cm}{\begin{flushleft}2
  j's~+~$\tau^+\,\tau^-$~+~$\slas{p}$; $l^+$~+~$\tau^+\,\tau^-$~+~$\slas{p}$.

(more  $\slas{p}$)\end{flushleft} }\\

 (e) $\left\{ \begin{array}{l} \sneut_\tau \rightarrow \tau^- \cplus \rightarrow
  \tau^- (f \bar f' \neut)\\ \stau_{1,2}^+ \rightarrow \tau^+ \, \neutdos
  \rightarrow \left[ \begin{array}{l} \tau^+ (Z^{(*)}) \neut)\rightarrow \tau^+
  (f \bar f) \neut \\ \mbox{ or } \\ \tau^+ (h^0 \neut) \rightarrow \tau^+ (b
  \bar b \mbox{ or } \tau^+ \tau^-) \neut \end{array}\right.  \end{array}
  \right.$& \parbox{7cm}{\begin{flushleft}4~j's~+~$\tau^+\,\tau^-$~+~$\slas{p}$;
  2~j's~+~$\tau^+\,\tau^-$~+~$l^+\,l^-$~+~$\slas{p}$;
  2~j's~+~$\tau^+\,\tau^-$~+~$l^+$~+~$\slas{p}$;
  $\tau^+\,\tau^-$~+~$l^+$~+~$l'^+\,l'^-$~+~$\slas{p}$.

($Z^0$ or $h^0$ emission;

more $b$ activity if $h^0$ or $Z^{*}\rightarrow b \bar b$; 

trilepton)\end{flushleft}}\\

 (f) $\left\{ \begin{array}{l} \sneut_\tau \rightarrow \tau^- \cplus \rightarrow
  \tau^- (f \bar f' \neut)\\ \stau_{2}^+ \rightarrow \bar\nu_\tau \cplus
  \rightarrow \bar\nu_\tau (f \bar f' \neut)\end{array}
  \right.$&\parbox{7cm}{\begin{flushleft} 4~j's~+~$\tau^-$~+~$\slas{p}$;
  2~j's~+~$\tau^-$~+~$l^+$~+~$\slas{p}$; $\tau^-$~+~$l^+\,l'^+$~+~$\slas{p}$.

(``wrong''-sign single $\tau$ lepton; 

same sign dilepton)\end{flushleft}}\\
\hline\end{tabular}
\caption{Detection signatures for the decay {\bf X}. The decay chains
  (a)-(c) are possible only if $m_{\sneut} > m_{\neut}$. The decay chains
  (d)-(f) are possible only if $m_{\sneut} >
  m_{\cplus}$. Here $\slas{p}$,j,$l^\pm$,$Z^{(*)}$ and $f$ denote missing
  energy-momentum, jet, isolated charged lepton, $Z^0$ real or virtual, and
  $(q,l^\pm,\nu)$, respectively.}
    \label{tab:canalstau}
  \end{center}
\normalsize
\end{table}

\end{document}